\documentclass[sigconf]{acmart}
\AtBeginDocument{%
  }

\copyrightyear{2026}
\acmYear{2026}
\setcopyright{cc}
\setcctype{by}
\acmConference[CHI '26]{Proceedings of the 2026 CHI Conference on Human Factors in Computing Systems}{April 13--17, 2026}{Barcelona, Spain}
\acmBooktitle{Proceedings of the 2026 CHI Conference on Human Factors in Computing Systems (CHI '26), April 13--17, 2026, Barcelona, Spain}
\acmDOI{10.1145/3772318.3790998}
\acmISBN{979-8-4007-2278-3/2026/04}

\usepackage{xcolor}
\usepackage{soul}
\usepackage{enumitem}
\usepackage{multirow}
\usepackage{graphicx} % Required for including images
\usepackage{subcaption} % Required for subfigures

\newcommand{\sysName}{BadminSense}

\begin{document}

\title{\sysName: Enabling Fine-Grained Badminton Stroke Evaluation on a Single Smartwatch}

\author{Taizhou Chen}
\orcid{0000-0002-7005-4560}
\affiliation{%
  \institution{Department of Computer Science, Shantou University}
  \city{Shantou}
  \country{China}}
\email{taizhou.chen@my.cityu.edu.hk}

\author{Kai Chen}
\orcid{0009-0002-4635-7462}
\authornote{The majority of this work was conducted while the author was at Shantou University}
\affiliation{%
  \institution{Department of Computer Science, Shantou University}
  \city{Shantou}
  \country{China}}
\affiliation{%
  \institution{CSSE, Shenzhen University}
  \city{Shenzhen}
  \country{China}}
\email{21kchen1@alumni.stu.edu.cn}

\author{Xingyu Liu}
\orcid{0009-0003-8625-3614}
\affiliation{%
  \institution{Department of Computer Science, Shantou University}
  \city{Shantou}
  \country{China}}
\email{21xyliu1@alumni.stu.edu.cn}

\author{Pingchuan Ke}
\orcid{0000-0003-2509-9046}
\affiliation{%
  \institution{Department of Sociology, Hong Kong Shue Yan University}
  \city{Hong Kong}
  \country{China}}
\email{pke@hksyu.edu}

\author{Zhida Sun}
\orcid{0000-0003-4689-986X}
\authornote{Corresponding author.}
\affiliation{
  \institution{CSSE, Shenzhen University}
  \city{Shenzhen}
  \country{China}
}
\email{zhida.sun@connect.ust.hk}

\newcommand{\revise}[1]{{\color{black}#1}}
\newcommand{\rerevise}[1]{{\color{red}#1}}

\renewcommand{\shortauthors}{Chen et al.}

\begin{abstract}
Evaluating badminton performance often requires expert coaching, which is rarely accessible for amateur players. We present \sysName, a smartwatch-based system for fine-grained badminton performance analysis using wearable sensing. Through interviews with experienced badminton players, we identified four system design requirements with three implementation insights that guide the development of \sysName. We then collected a badminton strokes dataset on 12 experienced badminton amateurs and annotated it with fine-grained labels, including stroke type, expert-assessed stroke rating, and shuttle impact location. Built on this dataset, \sysName~ segments and classifies strokes, predicts stroke quality, and estimates shuttle impact location using vibration signal from an off-the-shelf smartwatch. Our evaluations show that \sysName~ achieves a stroke classification accuracy of 91.43\%, an average quality rating error of 0.438, and an average impact location estimation error of 12.9\%. A real-world usability study further demonstrates \sysName’s potential to provide reliable and meaningful support for daily badminton practice.
\end{abstract}



\begin{teaserfigure}
  \includegraphics[width=\textwidth]{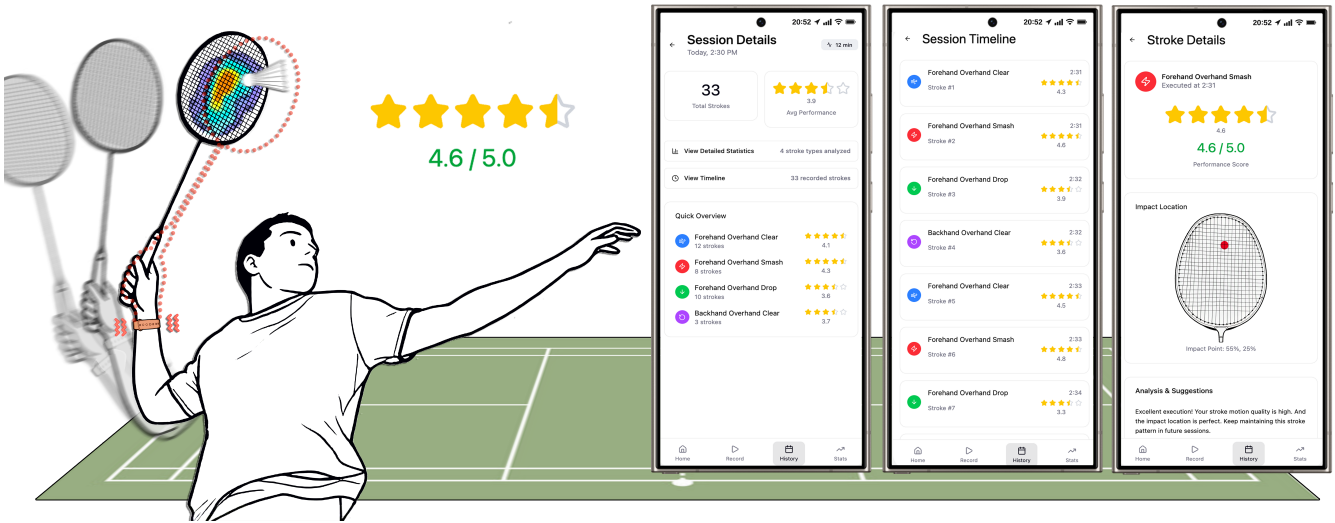}
  \caption{Illustration of the \sysName's~ concept. The system leverages a single smartwatch to detect and provide fine-grained badminton stroke analysis, including stroke classification, stroke quality rating, and impact location estimation. Three screenshots on the right illustrate the front-end interface of \sysName. Left: Session overview. This interface presents the stroke count in a game session with a performance summary. Middle: Session timeline view (\textbf{DR4}). This interface shows the sequence of strokes executed throughout the session period (\textbf{DR4}). Right: Stroke view. This interface displays the stroke quality rating (\textbf{DR1}), the impact location (\textbf{DR2}), and the improvement advice (\textbf{DR3}) from top to bottom.}
  \label{fig:teaser}
\end{teaserfigure}


\maketitle

\section{Introduction}

% % [real-time coaching is importance]
% Coaching plays a vital role in skill acquisition and performance improvement in sports\cite{elvitigala2024grand, nichol2019examining, jowett2017coaching}, especially for racket sports such as badminton\cite{hastie2009development, phomsoupha2015science, hambali2023coach}, tennis\cite{allison1980behavioral, reid2007skill}, and table tennis\cite{martinent2020literature, raab2005improving}, where performance relies heavily on precise swinging technique. 
% % [professional coaching is not accessible for daily practice]
% However, access to professional coaching in everyday practice is limited for amateurs.

% [Amateurs tend to improve their skill through daily practice]
Professional and amateur athletes are always looking for ways to improve their performance. A low-cost and accessible strategy to achieve this is deliberate practice, which plays a vital role in skill acquisition and performance improvement in sports\cite{ericsson1993role}, especially for racket sports such as badminton\cite{phomsoupha2015science}, tennis\cite{reid2007skill}, and table tennis\cite{zhang2012effects}, where performance relies heavily on precise swinging technique. 
% [It is hard to access to professional coaching in daily practice. Referring to coaching video is also less efficient]
However, access to professional coaching in everyday practice is limited for amateurs. Our preliminary interview reveals that amateurs tend to refer to online coaching videos and practice through imitation for skill improvement, but it appears to be less efficient without performance feedback\cite{arbabi2016effect}. 

% [rapid development of sensor provide technical supports]
In recent years, the rapid development of sensor technology has offered opportunities for the sport industry to develop sensor-based devices\cite{secckin2023review, aroganam2019review} aimed at activity tracking and performance analysis, thereby providing amateurs with accessible sensor-driven feedback during daily practice to facilitate deliberate practice\cite{krizkova2021sport}. 
% [powerful sensor solution on racket sport are relied on racket-mounted sensor or camera-based sensor system, which case inconvenient]
The racket sport community has also embraced the trend by introducing racket-mounted sensors, such as SmartDampener\cite{liu2024smartdampener}, Zepp\cite{zepp}, Babolat Play\cite{babolat_play}, Qlipp\cite{qlipp_tennis_qlipp_2017}, Sony Smart Tennis\cite{sony}, or camera-based systems, such as SwingVision\cite{swingvision}, Hawk-Eye\cite{hawk-eye}, and Playsight\cite{playsight}, to provide timely professional performance analysis. However, despite the racket-mounted sensor system and camera-based system being able to provide precise tracking and professional analysis results, these systems suffer from practical limitations. For example, racket-mounted sensors are mounted either on the string area\cite{liu2024smartdampener, qlipp_tennis_qlipp_2017} or on the handle\cite{zepp, babolat_play, sony}, which may affect the racket's vibration, weight, and balance\cite{park2024silent}. 
% Camera-based systems are considered invasive to privacy\cite{liu2024smartdampener} with low flexibility and accessibility since they often require external hardware setup\cite{park2024silent}. 
\revise{While camera-based systems are powerful, their effectiveness in badminton tracking is limited by factors such as the fast shuttle motion, variable lighting conditions, and the need for external equipment setup\cite{park2024silent}. }

% [wrist-worn device shows advantage on detecting arm movements]
In contrast to external sensor systems, wearable technology, such as smartwatches, offers more accessible and unobtrusive ubiquitous sensing opportunities for activity tracking\cite{turmo2021design, coorevits2016rise}. As a wrist-worn device, a smartwatch also shows great potential to serve as a vantage springboard to capture wrist and arm movements, which has attracted interest from both the academic community and the industry in exploring sensing techniques for racket sports. 
% [lack of fine-grained sensing solution for badminton on wrist-worn devices]
However, existing attempts that focus on measuring basic metrics, such as stroke classification\cite{park2024silent, anand2017wearable}, shot counting\cite{vivo, oppo}, and caloric measurement\cite{balance}. Compared with sensor-based and camera-based sensing systems, these approaches, while helpful for activity recording, lack support for precise and in-depth analysis of skill development in badminton\cite{elvitigala2024grand}. 

To address this gap, we propose \sysName, a smartwatch-based fine-grained badminton stroke sensing and skill analysis system to facilitate badminton self-training. \sysName~ exploits the built-in IMU sensor and microphone on a commercial smartwatch to capture both the motion sequence and the vibration from the user's wrist during a badminton stroke, and provides stroke-level performance skill analysis and evaluation. Inspired by our formative study on 5 badminton players, \sysName~ provides fine-grained badminton stroke detections and evaluations by jointly predicting stroke type, stroke quality, and shuttle impact location, and generates improvement advice based on these statistics. To this end, we collect a dataset consisting of 848 badminton strokes from 12 experienced badminton players, with each stroke annotated for both its shuttle impact location and quality by 21 advanced badminton players. We developed, trained, and evaluated \sysName~ on the dataset to enable stroke segmentation, stroke classification, stroke quality rating, and estimation of shuttle impact location. Our offline technical evaluation shows that \sysName~ achieves a user-independent stroke classification accuracy of 91.43\%, a user-independent average quality rating error of 0.438, and a user-independent average impact location estimation error of 12.9\%. In addition, a real-world usability study was further conducted on \sysName~ that highlighted its potential to provide unobtrusive and meaningful support for daily badminton training. 

% \sysName~ highlights the broader potential of wrist-worn devices to enable fine-grained sensing and analysis of racket sport, which could be further extended to other sports involving swinging motions, such as golf, tennis, and table tennis.

Our contributions are fourfold:
\begin{itemize}
\item We present a formative exploration that identifies requirements and guidelines for designing a badminton skill analysis and training system for a smartwatch.
\item We construct and open-source a dataset comprising 848 badminton strokes, each with expert-labeled action quality and impact location.
\item We develop a real-time system for analyzing badminton stroke-level performance skills and evaluation on an off-the-shelf smartwatch.
\item We conduct a usability study highlighting the insights and benefits of our proposed system.
\end{itemize}

\section{Related Works}
\sysName~ is largely inspired by existing works on motion analysis approaches for badminton and racket sports sensor systems. In addition, we also refer to the formative design method within the Sports Human-Computer Interaction (SportsHCI) community.  

\subsection{Badminton Performance Analysis}
Badminton performance analysis attracted attention from both academia and industry in the past decade. Existing literature explored the use of various sensing techniques, such as sensor-based approaches and computer vision-based approaches, for different purposes such as badminton shot detection\cite{oppo, vivo, rahmad2020vision, anand2017wearable}, badminton stroke classification\cite{zhu2025analysis, chang2025bst, wang2016badminton, steels2020badminton}, tactical analysis\cite{ibh2024stroke, weeratunga2017application, van2024strategy}, and education\cite{lin2023effect}. 

% Wearable sensor based
% single sensor
Detecting badminton stroke motion with a racket-mounted IMU sensor is a straightforward and effective approach. The sensor could be mounted either in the string area\cite{anik2016activity, towler2023effects} or in the handle area\cite{wang2019automatic, peralta2022badminton}. Specifically, Anik et al.\cite{anik2016activity} classified 3 types of stroke with an IMU sensor attached in the racket string area, while Wang et al.\cite{wang2019automatic} leveraged a deep-learning algorithm to process data from a racket-mounted IMU sensor to classify 10 types of badminton stroke. 
% multiple sensor
Researchers also explored the use of multiple motion sensors to support fine-grained badminton sensing\cite{wang2016badminton}. Steels et al.\cite{steels2020badminton} applied a data-driven approach to classify 7 types of badminton strokes with 3 IMU sensors mounted on the racket’s grip, the wrist, and the upper arm, respectively. 
% Similarly, Ma et al.\cite{ma2021retracted} detected badminton strokes with one racket-mounted IMU and two ankle-mounted IMU. 
Van et al.\cite{van2024strategy} proposed a system with a racket-mounted IMU sensor, a wrist-worn IMU sensor, and a body-worn UWB sensor to detect the player's stroke type and track their movements for tactical analysis. 
% wrist worn
Literature also investigated detecting badminton stroke from an arm- or wrist-worn device. Anand et al.\cite{anand2017wearable} pioneered the exploration of shot detection in swing sports, such as tennis, badminton, and golf, using a smartwatch-based sensing system. Lin et al.\cite{lin2023effect} measure the learner's forearm movement with a Myo armband to assist the coach for educational purposes.

% Computer Vision based
Compared with racket-mounted or body-mounted sensors, camera-based systems show the advantage of capturing players' movements from a global perspective. This advantage has attracted previous research to leverage computer vision-based approaches for tactical analysis\cite{weeratunga2017application, weeratunga2014application, chu2017badminton} and movement prediction\cite{ibh2024stroke, wang2023benchmarking}. Specifically, Weeratunga et al.\cite{weeratunga2014application, weeratunga2017application} processed fixed-view badminton match video footage for tactical classification and analysis. Using a similar data source, Ibh et al. \cite{ibh2024stroke} proposed a transformer-based encoder-decoder model to predict the next movement of the badminton athlete. Wang et al. \cite{wang2023benchmarking} contributed a dataset for badminton stroke landing points estimation.
\revise{
Prior researches also explores immersive visualization techniques for match-level tactical analysis \cite{chu2021tivee,lin2023vird} and shuttle trajectory analysis \cite{ye2020shuttlespace}. 
}

Besides, computer vision-based approaches have also been widely used for stroke classification\cite{chu2017badminton, chang2025bst, zhu2025analysis, rahmad2020vision}. Although camera-based sensing approaches often offer more precise tracking performance, they are considered invasive to privacy\cite{liu2024smartdampener} with low 
flexibility and accessibility as they often require external hardware setup\cite{park2024silent}. 

Despite the growing body of literature on wearable sensing technologies for sport in recent years, we observed that there is limited literature that focuses on badminton-specific applications. One possible reason is that, compared to other racket sports such as tennis and \revise{padel}, it remains a unique challenge as it is a fast-paced sport with a smaller hitting impact, which increases the difficulty of swing motion analysis using wearable devices. In addition, we also notice that the existing literature primarily focuses on stroke classification rather than evaluation, regardless of whether a computer vision-based approach or a sensor-based approach is employed. \sysName~ highlights the broader potential of wearable devices to enable fine-grained sensing and analysis of badminton sport. 

\subsection{Racket Sports Sensor Systems}

Badminton stroke shares a similar motion pattern with other racket sports such as tennis, table tennis, and padel games. Therefore, we also refer to the existing racket sports sensor system while designing \sysName. 
% Tennis, as one of the most popular racket sports around the world, have attracted attention from both academic community and industry to develop sensor system toward tennis motion tracking and analysis. 
Similar to badminton, approaches such as racket-mounted sensor\cite{liu2024smartdampener, blank2017ball, qlipp_tennis_qlipp_2017}, wrist-worn sensor\cite{park2024silent, zhao2019tenniseye, ganser2021classification}, and camera-based system\cite{gourgari2013thetis, swingvision, diogo2025smashing} have been explored towards the detection and analysis of racket sports motion. One of the straightforward and popular practices for detecting racket swing motion is using racket-mounted sensors. Commercial products for tennis analysis, such as Zepp\cite{zepp}, Babolat Play\cite{babolat_play}, Qlipp\cite{qlipp_tennis_qlipp_2017}, and Sony Smart Tennis\cite{sony} have achieved commercial success in the past decade. Exploration from academia also contributed significant insight to the community. SmartDampener\cite{liu2024smartdampener} integrated sensor techniques into a tennis dampener, mounted in the string area of the tennis racket. It enables ball speed estimation, impact location prediction, and stroke type detection by processing the vibration signal of the string area while stroking. TennisEye\cite{zhao2019tenniseye} contributed a physical model to estimate the ball speed with a racket-mounted IMU sensor. TennisMaster\cite{yang2017tennismaster} proposed a method for evaluating tennis serve performance through an IMU mounted on the user's shank and an IMU mounted on the racket. In addition to tennis, Blank et al.\cite{blank2015sensor} classified 8 types of table tennis strokes with a racket-mounted sensor. They further estimated table tennis ball speed and spin pattern using the same sensor setting\cite{blank2017ball}. 

A wrist-worn device also shows its advantage in detecting racket movement in other sports, especially those that involve significant arm swing movements\cite{jia2021swingnet}. Lopez et al.\cite{lopez2019site} proposed a system to measure baseball pitching action and tennis serve action using a smartwatch. Ganser et al.\cite{ganser2021classification} classified 5 tennis stroke types through a data-driven approach using a wrist-worn IMU sensor. Recently, Park et al. \cite{park2024silent} proposed Silent Impact, a real-time wearable system that detects and classifies six types of tennis strokes using a passive arm-worn smartwatch. In addition to motion data, Sharma et al. \cite{sharma2018wearable} fused acoustic data with motion data to detect racket shots on a smartwatch. 

Camera-based approaches are also widely explored to track and detect racket sport movement\cite{diogo2025smashing}. The SwingVision\cite{swingvision} and Playsight\cite{playsight} utilize a smartphone camera with computer vision techniques to provide racket motion statistics such as shot type, shot placement, ball speed, etc. However, such a system often requires an external camera setup and often suffers from environmental lighting conditions and occlusion problems. Numerous projects have also investigated capturing sports movement through 3D vision. Gourgari et al.\cite{gourgari2013thetis} contributed a 3D tennis motion dataset that included 12 types of tennis strokes. Research also utilizes a multi-camera system to collect fine-grained motion data, facilitating motion analysis\cite{skublewska-paszkowska_temporal_2023}. Such approaches offer high-fidelity tracking results, but require a sophisticated hardware setup, which is less practical for everyday use.

\subsection{Formative Exploration in SportsHCI}

Recent years have witnessed a growing body of research in sport Human-Computer Interaction (SportHCI)\cite{elvitigala2024grand} that aims to leverage interactive technology with human-centered design guidelines to enhance sport experiences for the public. \sysName~ also refers to those systems where formative exploration is widely adopted to facilitate the human-centered design process in their system design pipeline. Formative exploration methods, such as interviews\cite{weng2025bridging} and online surveys\cite{lin2022quest}, have been employed in recent SportHCI research. Specifically, Ma et al.\cite{ma2024avattar} conducted interviews with 11 table tennis players to elicit four key design requirements and developed avaTTAR, an AR-based stroke training system. Similarly, VisCourt\cite{cheng2024viscourt} proposed a mix-reality basketball tactical training system through formative interviews with both coaches and professional players. iBall\cite{zhu2023iball} focused on augmenting basketball game viewing experiences through formative studies with basketball fans. In addition to pure interviews, literature also combines interviews with online surveys to obtain more comprehensive information. PoseCoach\cite{liu2022posecoach} combined an online survey and expert interviews to inform the design of a video-based running coaching tool. Similarly, Wu et al.\cite{wu2023ar} applied formative insights from fitness enthusiasts to design an AR-guided at-home workout interface.

\sysName~ follows the user-centered design guideline where we polish the system design requirements through a formative exploration with six die-hard badminton amateurs.

\section{Formative Exploration}

Our exploration of \sysName~ begins with a formative study aimed at formulating the design requirements of \sysName~ through a deep understanding of our target users in the context of badminton training. To this end, we conducted a semi-structured interview with experienced amateur players to understand their gap\revise{s} and challenges encounter\revise{ed} during self-training, their perspectives \revise{on} badminton stroke skill, and their expectation\revise{s} of a wearable training assistant system. Accordingly, we raise the following three research questions to guide our exploration on the design of \sysName:

\begin{enumerate}
[itemsep=.5em, label={\textbf{RQ\arabic*}}]
    \item How do experienced players currently evaluate and improve their performance in the absence of professional coaching?

    \item What are experienced players' expectations and concerns regarding \revise{} technology designed for detailed badminton performance analysis?

    \item What fine-grained factors of a badminton stroke do they consider critical to its performance, and how do they evaluate them?
\end{enumerate}

\revise{\subsection{Form Factor Selection}

\sysName~ is highly motivated by the insights and suggestions from existing studies on SportHCI \cite{elvitigala2024grand} and sport wearable \cite{secckin2023review}. First, Elvitigala et al. \cite{elvitigala2024grand} emphasize the importance of a sportHCI system to support robust and reliable sensing for longitudinal use. Modern smartwatches employ technically mature sensor integration, which can offer stable sensing performance over time. Second, literature \cite{elvitigala2024grand} also calls for unobtrusive technologies that minimize disruptions to athletes' experience. Smartwatches are lightweight, familiar to users, and generally more socially acceptable than alternative form factors such as smart wristbands or camera-based systems. Last but not least, Elvitigala et al. \cite{elvitigala2024grand} further highlight that cost and accessibility are key considerations for real-world adoption. A survey \cite{secckin2023review} on sport wearables further reveals that the smartwatch is the most popular form factor among all wrist-worn devices (10/13 products). Thus, we focus our exploration on the smartwatch form factor. }

\subsection{Expert Interview}

\subsubsection{Participants}

We recruited five right-handed male badminton players aged 20 to 23 ($M = 22$, $SD = 1.22$) through university sports clubs and social media groups, with a minimum requirement of six years of consistent play. Their self-reported average badminton experiences were 7.8 years ($SD = 1.10$). All participants were non-sport majors with diverse backgrounds\revise{,} including Computer Science, Law, Mathematics, and Civil Engineering. We consider that this subject group is well-suit\revise{ed} for our study. On one hand, they are highly motivated to improve their skills but often lack the resources or time to access professional coaching, leading to rich self-training experiences. On the other hand, their comprehensive understanding of badminton techniques enabled them to provide insightful comments and feedback for our system design. All participants had experience of using \revise{a} smartwatch for fitness tracking, including monitoring heart rate and measuring caloric consumption. \revise{The study was reviewed and approved by the institutional Human Research Ethics Committee. All participants were provided informed consent before participation, and they received a 5 USD coupon for compensation. } 

\subsubsection{Interview Topics}

To answer our research questions, our semi-structured interview involved several key topics, including:

\begin{enumerate}[label={}, leftmargin=*]

    \item \textbf{Badminton Training Experience:} Participants were invited to share their personal experiences with badminton stroke skill training, including their experiences and expectations of professional coaching, and their experiences and methods of self-training. 
    \item \textbf{Gap and Challenges:} We particularly asked them to identify the gap\revise{s} and challenges they encounter during self-training.
    \item \textbf{Expectation of Technology:} As all participants had prior experience of using smartwatches for fitness tracking, we particularly encouraged them to \revise{share} their concerns, suggestions, and envision their ideal form of a smart wearable device for badminton training and performance training. 
    \item \textbf{Factors Affecting the Quality of Badminton Stroke:} We elicited the participants to discuss their knowledge of key factors that influence the quality of a badminton stroke.
\end{enumerate}

\subsection{Findings and Insights}

We gained valuable findings and insights during our interview (\textbf{F\#}). These insights helped us understand the players' self-training behaviors and pain points that they encounter when training alone, and key aspects of performing badminton strokes. These insights not only assist our system design but also inspire our system implementation.

\begin{enumerate}
[itemsep=.5em, label={\textbf{F\arabic*}}]

% RQ1 experiences
% RQ2 expectations
% RQ3 factors

\item \textbf{Deliberate Practice through Online Resources.}
Participants reported that much of their self-training relies on external resources such as online coaching videos and match recordings, with practice often based on imitation. Three participants explicitly mention\revise{ed} of engaging \revise{in} deliberate practice specifically for a certain technique, such as strengthening particular muscle groups, or familiarizing themselves with the correct impact location through ball-feeding practice (\textbf{RQ1}).

\item \textbf{Lack of Feedback without a Coach.}
Since most daily practice occurs without a coach, self-training with self-assessment becomes essential. A major challenge participants identified is the difficulty of determining whether a stroke is executed correctly. A large proportion of participants have experience \revise{in} recording themselves with video and playback afterward for self-assessment. However, they are still eager for quantitative and professional feedback on their performance (\textbf{RQ2}).

\item \textbf{Need for Longitudinal Performance Tracking.}
Participants also noted that their actual movements tended to be deformed during long-term play, resulting in performance bias compared with their deliberate practice. Therefore, they explicitly mention the need \revise{to support} longitudinal quantitative performance tracking to guide and adjust their training plans (\textbf{RQ2}). 

\item \textbf{Factors Affecting Stroke Performance.}
All participants agree \revise{that} the speed and the direction of the shuttle are crucial factors to a stroke performance, and that both stroke quality and the impact location significantly affect the control of these factors. Specifically, a hit in the ``sweet spot'' of the racket string area would maximize the speed and strength for offensive strokes such as \textit{Smash} and \textit{Clear}. In contrast, striking closer to the racket’s edge provides easier control for defensive strokes such as \textit{Drop} through spinning or eliminating incoming strength. Three participants further emphasized \revise{that} the correctness of the stroke motion pattern would guarantee the efficient transfer of muscle power, making the stroke more controllable and powerful. Additionally, one participant highlighted the importance of footwork, comment\revise{ing} that proper footwork allows the player to hit the ball with the best timing and force (\textbf{RQ3}). 

\item \textbf{Self-Assessment through Racket and shuttle Feedback.}
Participants often judge their stroke quality based on the impact sound, haptic sensations, and the shuttle flight trajectory. For example, a crisp sound is often taken as a sign of hitting the ``sweet spot'', often accompanied by a significant and prolonged racket vibration. However, the correctness of this judgment, which is usually subjective, is largely dependent on player experiences (\textbf{RQ3}).

% \item \textbf{Openness to Wearable Technology.}
% While two of the participants preferred wearing a smartwatch on the non-dominant hand to monitor physiological indicators during play badminton play, all participants show willing to wear it on the dominant hand if it could track and provide fine-grained and useful feedback of their performance. The tracking experience should be unobtrusive and require minimum interaction.

\end{enumerate}

\subsection{System Design Requirements}

Based on our empirical findings and insights \revise{from} the interview, we distilled four system design requirements (\textbf{DR\#}) of \sysName, together with three implementation insights (\textit{I\#}) that guide our system pipeline and algorithm design.

% Stroke Quality Evaluation(F2)
% # stroke-specific evaluation, classification before evaluation(F4)

% Impact Location(F4)
% # using sound and vibroation(F5)

% Feedback for improvment based on the quality and impact location(F3)

% Longitudinal Post-Session Analytics(F3, F6)

\begin{enumerate}
[label={\textbf{DR\arabic*}}]

\item \textbf{Stroke Motion Quality Evaluation.}
Participants emphasized the importance of stroke motion quality (\textbf{F4}) and highlighted the difficulty of judging whether a stroke is performed correctly without \revise{a} coach (\textbf{F2}). Therefore, the system should support stroke motion quality evaluation by providing players with quantitative assessments of their stroking skill. 
\begin{enumerate}
[label=\textit{I1:}]
\item To ensure fair and accurate assessment, stroke-specific evaluation strategies are needed to align the motion difference in stroking skill. The system should classify the stroke types (e.g., Smash, Clear, Drop) before applying the evaluation algorithm (\textbf{F4}).
\end{enumerate}

\item \textbf{Impact Location Estimation.}
As \revise{the} impact location being pointed out \revise{is} a crucial factor affecting stroke quality, the system should be able to estimate the shuttle impact location on the racket string area. This enables fine-grained stroke evaluation and skill suggestion, such as distinguishing between sweet-spot hits for offensive strokes and edge-area hits for defensive strokes (\textbf{F4}).
\begin{enumerate}
[label=\textit{I2:}]
\item Acoustic and vibration signals produced at the moment of impact may serve as \revise{an} effective feature for estimating the impact locations (\textbf{F5}). 
\end{enumerate}

\item \textbf{Feedback for Improvement.}
To compensate for the absence of a coach, the system should provide not only quantitative feedback but also qualitative feedback that is more user-friendly for stroke improvement (\textbf{F2}). 
\begin{enumerate}
[label=\textit{I3:}]
\item Since each stroke type depends on different skill criteria, feedback should jointly consider stroke type, stroke quality, and the impact location data to provide precise and fine-grained guidance (\textbf{F4}).
\end{enumerate}

\item \textbf{Longitudinal Analytics}
The system should offer post-session performance visualization and analytics to help players identify their skill gaps and track their progress across daily practice through session historical playback. The system should work unobtrusively \revise{minimal} interaction required (\textbf{F3}).

\end{enumerate}

\subsection{Post-Session Interview}

\textbf{DR3} with \textit{I3} highlighted the need to identify concrete skill criteria for different stroke types, especially with respect to the impact location. To the best of our knowledge, previous research and documentation pay limited attention \revise{to} discussing the effect of impact location on other stroke type\revise{s} apart from smash\cite{mcerlain2020effect, towler2023effects}. To support the proof-of-concept development of \sysName, we conducted a short post-session interview \revise{with} six expert players to identify stroke-specific skill criteria focusing on the impact location across four representative strokes.

\begin{figure*}[h!]
    \centering
    \includegraphics[width=\linewidth]{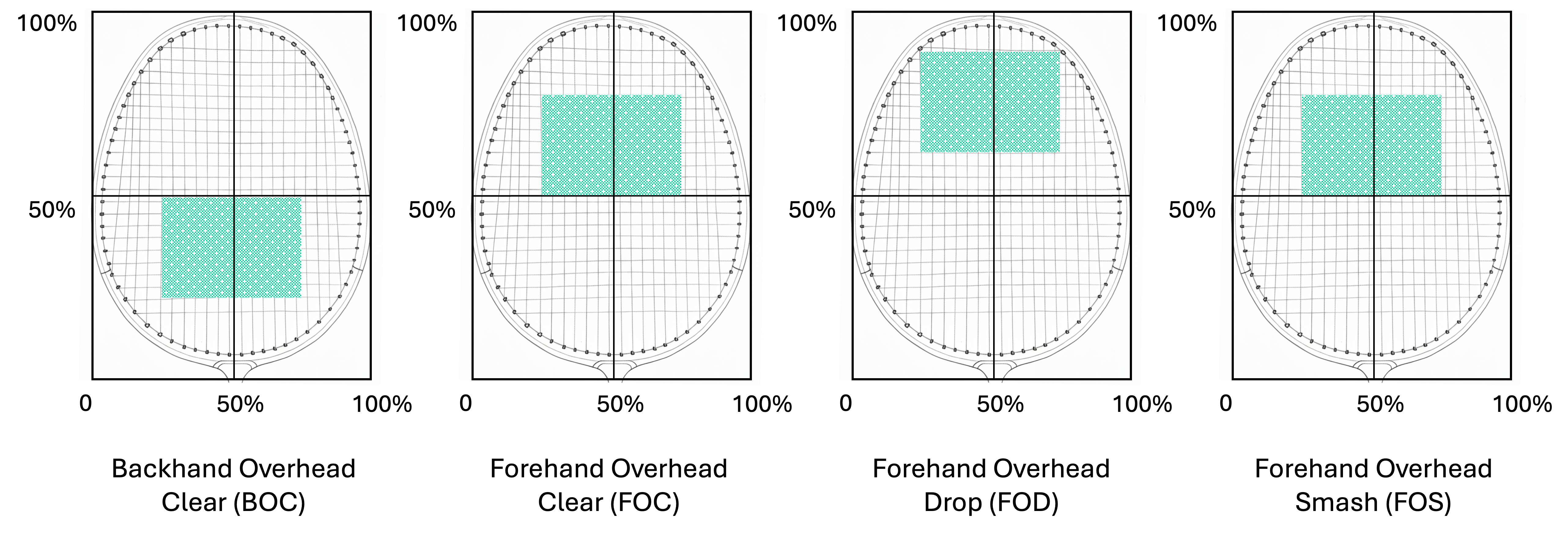}
    \caption{Stroke-specific impact location that most users prefer. The light green area indicate\revise{s} the optimal impact location.}
    \label{fig:optimal_impact_location} 
\end{figure*}

\subsubsection{Stroke Selection}

Our study focuses on four basic and representative badminton strokes: \textit{forehand overhead clear}, \textit{forehand overhead smash}, \textit{forehand overhead drop}, \textit{backhand overhead clear}. The first three forehand strokes are recognized as fundamental techniques in badminton training\cite{grice_badminton_2008} and frequently occur during gameplay\cite{laffaye2015changes}. While backhand strokes appear less frequently in badminton games\cite{phomsoupha2015science}, prior studies have also shown that common backhand strokes share similar motion patterns\cite{huang2002kinematic}. Hence, we selected the \textit{backhand overhead clear} as a representative stroke in this proof-of-concept investigation. 

\subsubsection{Participant, Task, and Procedure}

We required six expert player\revise{s} (age: $M = 28.5$, $SD = 2.89$) through \revise{the} university sports club, including three die-hard badminton amateur players and three university badminton coaches. During the study, participants were provided a tablet displaying an image of a badminton string area. \revise{They} are required to mark the impact location that they considered optimal for each of the four strokes using a stylus. They were then invited to explain the reasoning behind their markings. The study lasted less than ten minutes.

\subsubsection{Results}

All participants agree that for \textit{forehand overhead clear} and \textit{forehand overhead smash}, the best impact location was in the upper-middle region of the racket face, between 50\% and 75\% vertically, and within the horizontal range of 25\% to 75\%. This finding also aligns with previous \revise{research} on \revise{the} effect of badminton impact location \cite{mcerlain2020effect, towler2023effects}. 

For the other two strokes, participants expressed more diverse views. For the \textit{backhand overhead clear}, five participants agreed that the optimal location was in the lower-middle region (25\%–50\% vertically, 25\%–75\% horizontally), while one \revise{participant} held a different view and commented that the optimal impact position was in the middle (25\%–75\% vertically, 25\%–75\% horizontally) of the racket face. For the \textit{forehand overhead drop}, four participants believed the optimal location should be slightly higher than that of the \textit{forehand overhead smash} (70\%–90\% vertically, 25\%–75\% horizontally). One participant suggested it should be the same as the \textit{forehand overhead smash}, and one participant argued for a lower-middle location (25\%–50\% vertically, 25\%–75\% horizontally). Figure. \ref{fig:optimal_impact_location} \revise{summarizes} the stroke-specific optimal impact location of each stroke.

\section{Dataset Acquisition}

\begin{figure*}[h]
    \centering
    \includegraphics[width=\linewidth]{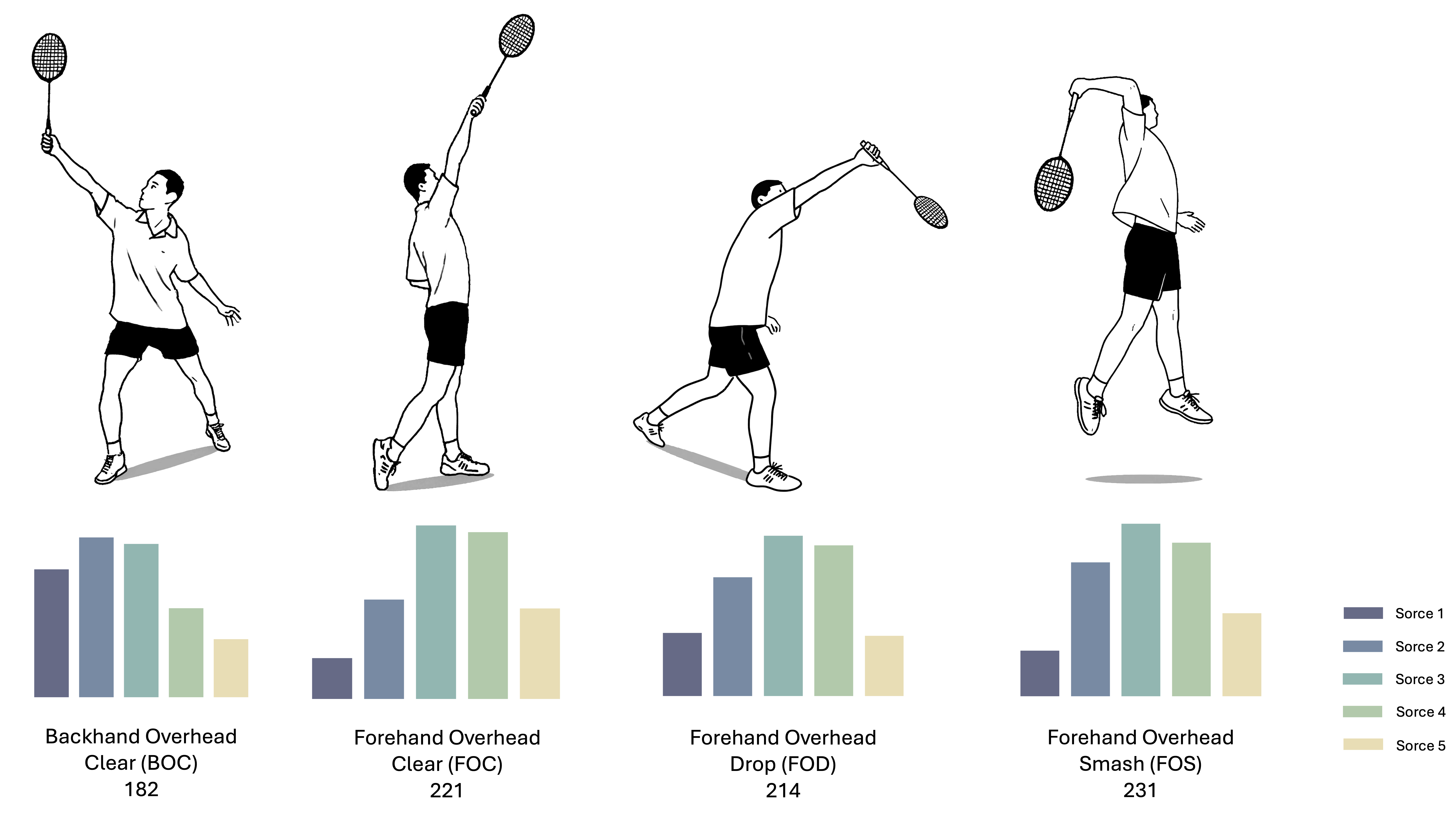}
    \caption{Illustration of the four stroke types that we focused on in this study. The corresponding bar chart below visualizes the distribution of the number of ratings, with the digit below indicating the total sample size in the category.}
    \label{fig:dataset_rating} 
\end{figure*}

To train and evaluate the proposed algorithm, we constructed a dataset on 12 amateur badminton players from a local university, representing diverse ages (Mean: 23.58, SD: 1.04), genders (10 males, 2 females), and skill levels (5 beginners, 5 intermediate, 2 advanced). \revise{Participants were recruited from \revise{the} institutional badminton club through word of mouth.} All of them are right-handed. Following Liu et al.\cite{liu2024smartdampener}, we categorized player expertise based on self-reported experience: beginner players had played badminton for less than 2 years, intermediate players for 3–5 years, and advanced players for more than 5 years. To ensure participants had sufficient skill with the sport to perform the required strokes, we recruited participants with a minimum requirement of badminton experience for 6 months, as recommended by Park et al.\cite{park2024silent}. Our data collection study focuses on four basic badminton strokes: \textit{forehand overhead clear}, \textit{forehand overhead smash}, \textit{forehand overhead drop}, \textit{backhand overhead clear} (Figure. \ref{fig:dataset_rating}). The first three forehand strokes are recognized as fundamental techniques in badminton training\cite{grice_badminton_2008} and frequently occur during gameplay\cite{laffaye2015changes}. While backhand strokes appear less frequently in badminton games\cite{phomsoupha2015science}, prior studies have also shown that common backhand strokes share similar motion patterns\cite{huang2002kinematic}. Hence, we selected the \textit{backhand overhead clear} as a representative stroke in this proof-of-concept investigation. 

Recall that \sysName~ is focusing on detecting, recognizing, and evaluating badminton strokes. To this end, we aim to assess stroke quality rating and estimate the shuttle impact position. To achieve this, we collected the motion and acoustic data using a dominant hand-worn smartwatch from every participant, with each stroke annotated with its stroke type, quality rating, and shuttle impact location. The video data \revise{were} recorded from a smartphone concurrently with the sensor data for data alignment and labeling. We \revise{open-source} the dataset to the community for further investigation and development\footnote{https://github.com/taizhouchen/BadminSense\_Dataset}.

\begin{figure*}[htb]
    \centering
    \includegraphics[width=\linewidth]{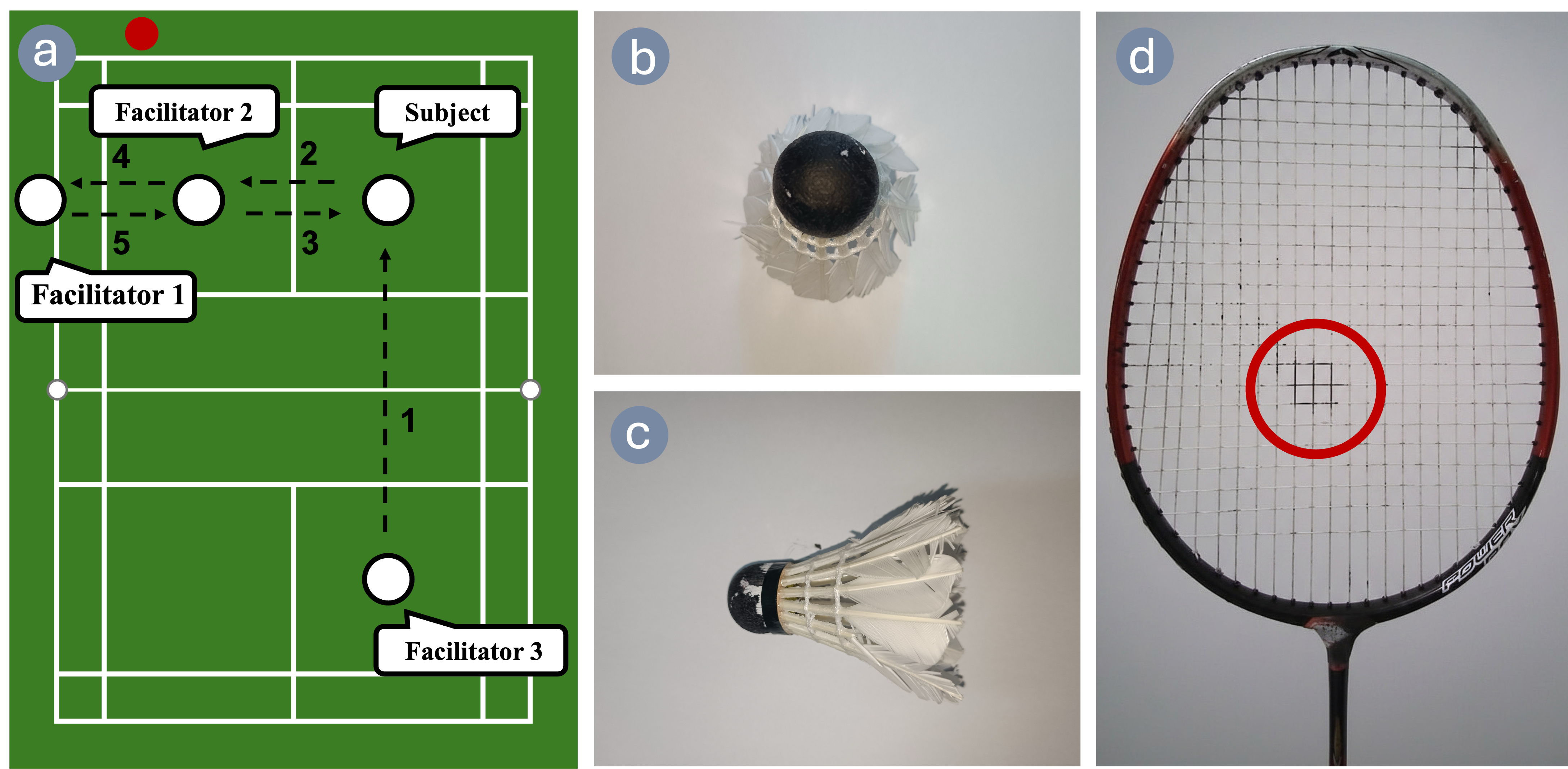}
    \caption{(a) Configuration of the badminton court for data collection and the data collection procedure. 1. Facilitator 1 serves a shuttle sprayed with ink, 2. Subject returns the racket after stroking, 3. Facilitator delivers a cleaned racket for the next stroke, 4. Facilitator 1 records the ink mark and cleans up the racket, 5. Facilitator 2 prepares a clean racket for the next stroke. The red dot \revise{in} the top left corner denoted the smartphone camera recording position for ground truth labeling. (b)(c) Shuttle with fountain pen ink on the head before serving. (d) An ink mark on the racket face by the shuttle impact after stroking, highlighted in a red circle.}
    \label{fig:datacollection} 
\end{figure*}

\subsection{Apparatus}

Our customized data collection system consists of a front-end smartwatch (Samsung Galaxy Watch 6 FE) for capturing motion and acoustic data, a front-end smartphone (Redmi Note 10 Pro) for capturing video data, and a back-end Windows server (11th Gen Intel(R) Core(TM) i5-11300H) for data storage and service control. A customized WearOS program and an Android program \revise{were} deployed on the smartwatch and the smartphone, respectively, for data streaming. The WearOS program was developed based on V Mollyn et al.'s implementation \cite{mollyn2022samosa}. The smartwatch streams IMU at 100Hz and microphone signals at 16,000Hz to a Python-based server via the TCP/IP protocol, while the smartphone simultaneously streams video data through the same protocol. To ensure temporal alignment across data frames stream from multiple clients, we implemented a heartbeat-based network synchronization mechanism. The badminton rackets that we used for the study are with string tensions of 24lbs\revise{,} which is suitable for both beginners and experts. During the study, we provide\revise{d} two Samsung Galaxy Watch 6 in different sizes (42mm and 48mm) and allow\revise{ed} the participants to \revise{choose} one that best fit their wrist. 

\subsection{Task and Procedure}

Our data collection study was carried out on a standard badminton court, with each session involving three experiment facilitators and one participant. One facilitator (Facilitator 3 in Figure. \ref{fig:datacollection}(a)), a professional badminton athlete, was responsible for serving the shuttle and instructing the participant in executing the specified strokes. In order to record the impart location, we implemented a novel data collection approach inspired by McErlain-Naylor et al.\cite{mcerlain2020effect} and Liu et al.\cite{liu2024smartdampener}. Before each serve, the shuttle head was sprayed with fountain pen ink (Figure. \ref{fig:datacollection}(b)(c)), allowing the point of contact on the racket string area to be visibly marked. Following each stroke, the racket with an inked impact location on its face (Figure. \ref{fig:datacollection}(d)) was photographed with a USB camera for impact location ground truth labeling. The racket string was cleaned up after each stroke, and the shuttle was replaced after 20 strokes to prevent weight gain due to ink accumulation. To facilitate this process, three identical badminton rackets were used in turn for each participant. Please refer to Figure. \ref{fig:datacollection}(a) for the detailed procedure. 

Upon the arrival of a participant, the facilitator introduced the study purpose and asked the participant to fill out the pre-study questionnaire for his/her anonymous biographic information, and sign the consent form voluntarily. Each participant was required to repeat each stroke 20 times with our smartwatch on \revise{their} dominant hand, yielding 4 $\times$ 20 = 80 strokes from one participant. To preserve signal diversity and reduce sensor bias, participants are instructed to take off the smartwatch and re-wear it after every 10 strokes. The order of stroke type was presented in the Latin-square-based counterbalanced order across all the participants to mitigate ordering effects. \revise{The study was reviewed and approved by the institutional Human Research Ethics Committee. All participants were provided informed consent prior to participation.} Each session lasted approximately one and a half hours, and participants were compensated with 7 USD for their time.

\subsection{Data Labeling}

\subsubsection{Segmentation and Impact Location}

We developed a graphical user interface with Python to facilitate the data labeling task. Each stroke was manually segmented from the IMU and acoustic data sequence based on the corresponding video recordings. We removed \revise{the} data that \revise{was} corrupted due to hardware or connectivity issues, as well as those that could not be labeled. As a result, we obtained 848 valid stroke samples, each consisting of an IMU motion sequence, a corresponding acoustic signal segment, and a corresponding video clip. For each stroke, the impact location was manually annotated, referring to the ink mark left on the racket string area on the captured image. \revise{Furthermore, to increase the variability of our training data, we applied standard data augmentation techniques to the valid stroke samples, including random temporal shifting, temporal scaling, and additive noise, across all IMU signal channels in the temporal domain. These operations expand the dataset by three times, to 2544 in total.}

\subsubsection{Stroke Quality Rating}
\label{SEC:stroke_quality_rating}

Next, we recruited 21 advanced badminton players with a minimum badminton experience requirement of 5 years to rate each stroke on a five-point Likert scale. \revise{The participants were recruited from the institutional badminton club forum.} To mitigate the intra-subject agreement bias, we provide reference video clips\footnote{Forehand Overhead Clear: https://youtu.be/S2brZPqx288}\footnote{Forehand Overhead Smash: https://youtu.be/HS3x2lX0Uao}\footnote{Forehand Overhead Drop: https://youtu.be/31O\_WuhVbKw}\footnote{Backhand Overhead Clear: https://youtu.be/wlCkx5R3gww} representing a stroke rated at the highest score of 5 for each stroke type. To minimize subject fatigue and ensure rating diversity, we randomly divided the dataset into three equal-sized folds, assigning each participant to one of them. Consequently, each stroke was rated by seven independent assessors. \revise{All participants were compensated with 14 USD for their time.}

A web-based rating interface was developed to facilitate the rating process. All assigned strokes were presented by their video clip in a randomized order. Assessors could preview the video by hovering their cursor over the video thumbnail, with a corresponding reference video clip appearing in a pop-up window concurrently. They were instructed to rate each stroke using a numerical rating slider based on their perception of stroke quality. 

To evaluate the reliability of the rating between all assessors, we calculated the Intraclass Correlation Coefficient (ICC)\cite{koo2016guideline} on each dataset fold based on three mean ratings (k = 7), consistency, and 2-way mixed-effects models. The results indicated good reliability for fold 1 (ICC = 0.824, 95\% Confidence Interval = [0.79, 0.85]), fold 2 (ICC = 0.865, 95\% Confidence Interval = [0.84, 0.89]), and fold 3 (ICC = 0.790, 95\% Confidence Interval = [0.75, 0.82]). In summary, the results suggest that the collected rating scores maintain a strong level of intra-subject agreement.

\section{BadminSense}
\subsection{System Overview}

\begin{figure*}[h]
    \centering
    \includegraphics[width=\linewidth]{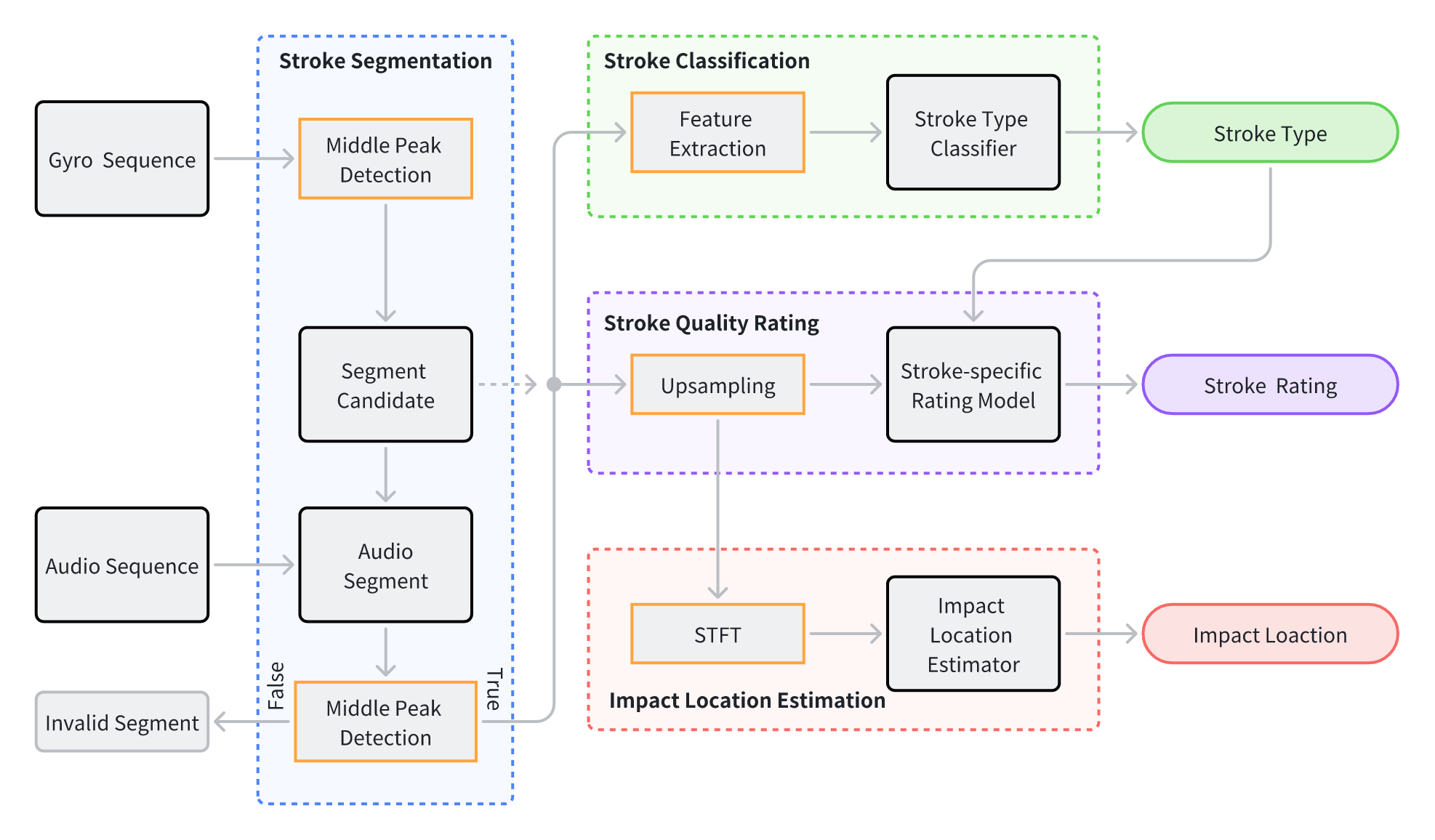}
    \caption{\sysName~ system pipeline overview. The pipeline consists of four components: stroke segmentation, stroke classification, stroke quality rating, and impact location estimation. }
    \label{fig:pipeline} 
\end{figure*}

We designed and implemented \sysName's~ system pipeline (Figure. \ref{fig:pipeline}) to fulfill the aforementioned system design requirements. The pipeline consists of four core components: stroke segmentation, stroke classification, stroke quality rating, and impact location estimation. In this section, we discuss four components and their evaluation results in detail.

\subsection{Stroke Segmentation}

Upon recording a new badminton session signal sequence that could consist of multiple stroke events, \sysName~ first detects, segments, and classifies each of them out of the whole sequence for further analysis. To reduce computational cost while preserving segmentation accuracy, we adopt a two-step stroke segmentation strategy that leverages both the IMU sequence and the audio sequence. 

\subsubsection{Approach}
\begin{figure*}[htb]
    \centering
    \includegraphics[width=\linewidth]{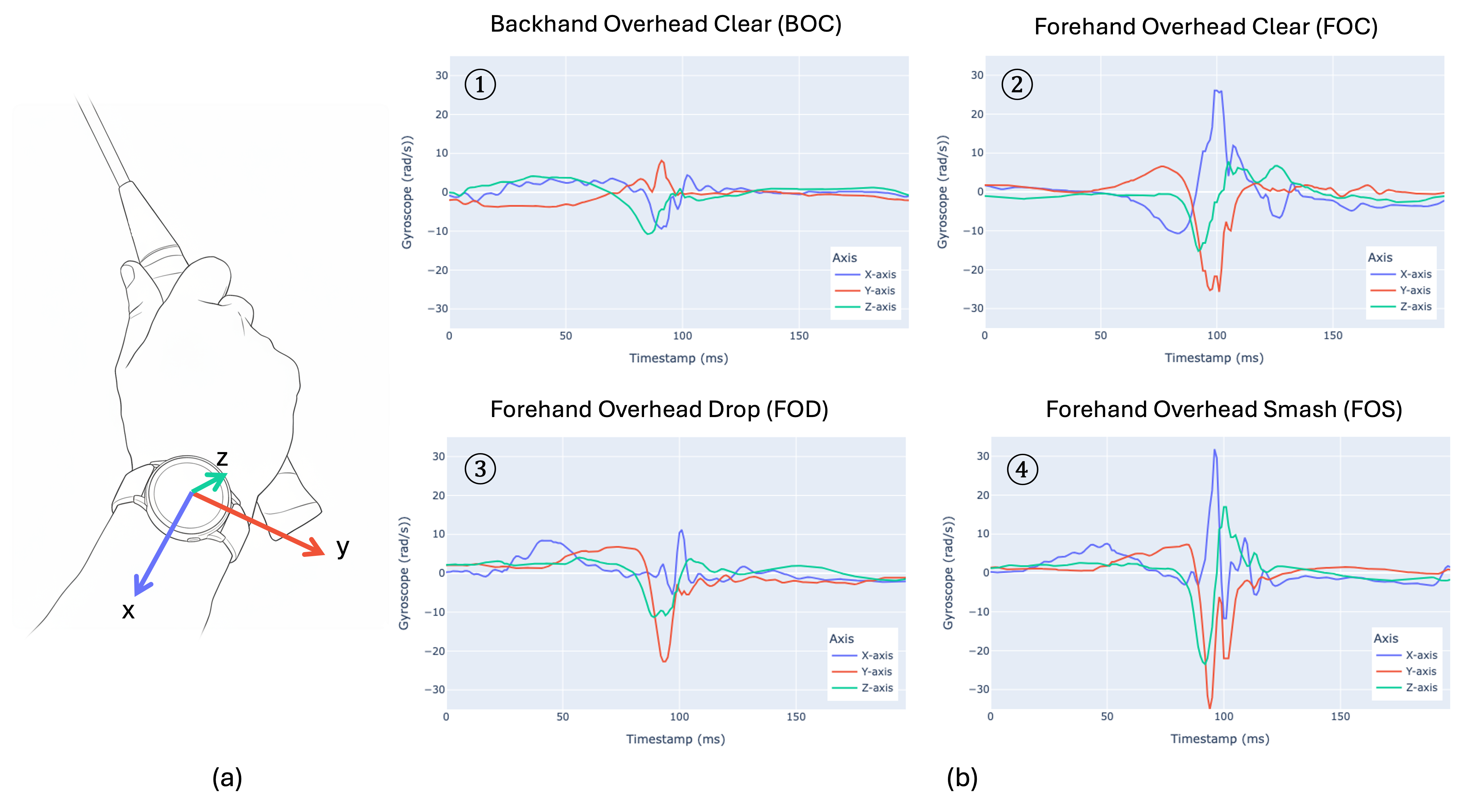}
    \caption{(a) Illustration of the smartwatch coordinates. (b) Gyro data sample of each stroke from one user.}
    \label{fig:coordinate} 
\end{figure*}

\begin{figure*}[h] % [h!] is a placement specifier (here, try to place it here)
    \centering % Centers the entire figure
    \begin{subfigure}[b]{0.75\textwidth} % [b] for bottom alignment, 0.45\textwidth sets width
        \centering
        \includegraphics[width=\linewidth]{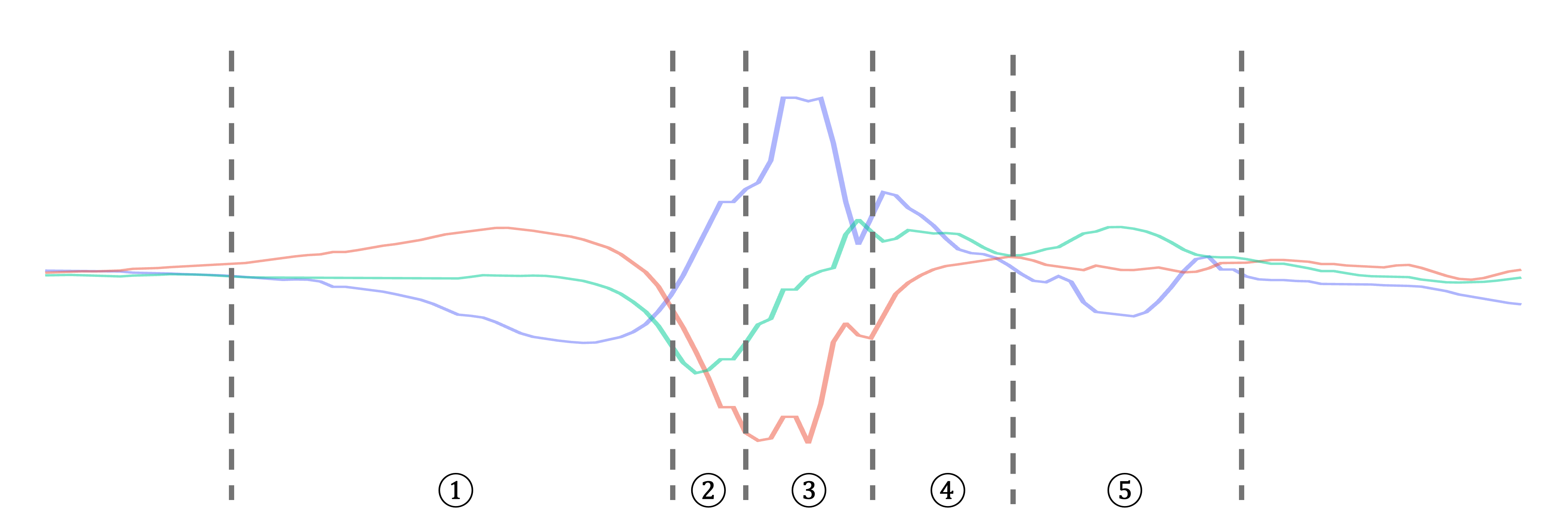} % Include your image
        \caption{Signal sample with phase segmentation of a forehand overhead clear stroke.}
        \label{fig:phase_signal_segment}
    \end{subfigure}
    \hfill % Adds horizontal space between subfigures on the same line
    \begin{subfigure}[b]{0.75\textwidth}
        \centering
        \includegraphics[width=0.9\linewidth]{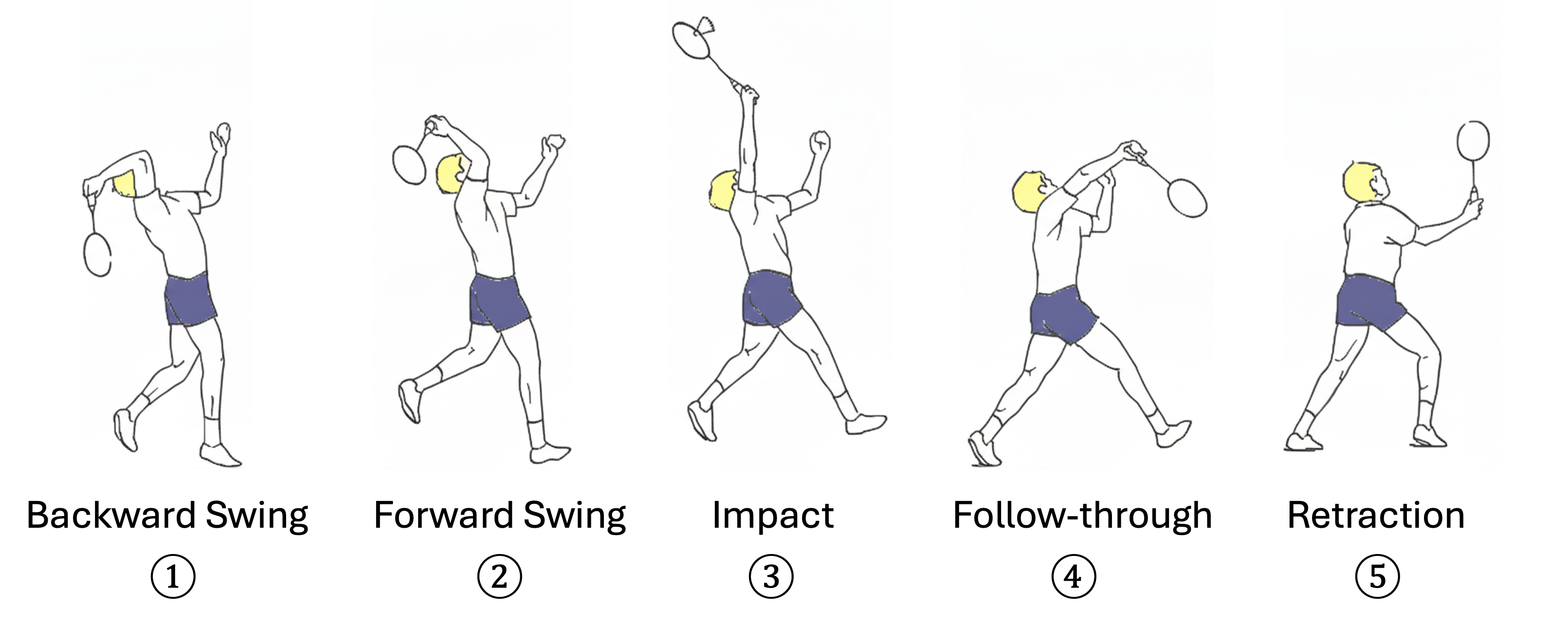}
        \caption{Illustration of stroke motion sequence phases. A complete stroke motion sequence generally comprises five sequential phases: backward swing, forward swing, impact, follow-through, \revise{and} retraction.}
        \label{fig:phase_motion_segment}
    \end{subfigure}
    \caption{Pair-wise illustration of five sequential phases on a \textit{forehand overhead clear} stroke sample. The five-phase segments in (a) are produced with a corresponding phase motion as illustrated in (b).}
    \label{fig:sequence}
\end{figure*}

We first exploited a \revise{lightweight} sliding window-based algorithm on the IMU sequence to detect a region of interest that potentially contains a stroke event.
% We observed that a stroke event occurs always accomplished by an instantaneous significant wrist rotational movement along the y axis (Fig. \ref{fig:coordinate}(a)). 
We observed that a stroke event always occurs by an instantaneous significant wrist rotational movement along the y-axis (Figure. \ref{fig:coordinate}(a)). Motivated by this, we implemented a peak detection algorithm on the y-axis gyroscope signal from the IMU. As Anand et al.\cite{anand2017wearable} pointed out, a complete stroke motion sequence generally comprises five sequential phases: backward swing, forward swing, impact, follow-through, and retraction. Each of these phases generates signal impulses, with the impact phase, typically occurring near the middle timestep, producing the most significant peak (see Figure. \ref{fig:phase_signal_segment}\textcircled{3}). Therefore, we treat a stroke event candidate as a window whose local maximum is located in the middle. This process yields a set of candidate windows represented by $\{ (t_i^{\text{start}}, t_i^{\text{end}}) \}_{i=1}^{N}$, where each pair represents the start and end time step of the $i^{th}$ window, and there were $N$ sets in total. We empirically set the window size to 2000ms and squared the sensor values before applying a local maximum threshold, which was empirically set to 21.
% with a local maximum threshold at 21. 
Moreover, we observed that the signal peak occurs slightly earlier than the impact point. Therefore, we empirically set a window offset of 100ms to locate the actual stroking event more precisely.

During practice, we observed that those racket swing movements without hitting the shuttle would share a pattern similar to that of a normal stroke, which increases the false positive (FP) rate of our segmentation algorithm. Note that a successful shuttle \revise{hit} would also produce significant impact sound. Inspired by this, we adopt a two-step verification utilizing the audio signal sequence. For each pair of time steps in $\{ (t_i^{\text{start}}, t_i^{\text{end}}) \}_{i=1}^{N}$, we extract the corresponding time windows segmented on the audio signal energy sequence. A similar peak detection algorithm was applied to the audio window. We only accept those windows that appear with a middle-located local maximum peak on both the IMU signal sequence and the audio signal sequence as valid stroke event segments.

\subsubsection{Evaluation}

We evaluate the aforementioned two-step segmentation approaches on our dataset. To eliminate the time-step bias during labeling, we allowed an error tolerance threshold of 200ms during evaluation. The results show that our approach achieves an average stroke segmentation accuracy of 99.41\% with a false positive rate of 0.23\%.

\subsection{Stroke Classification}\label{SEC:classification}

\textit{I1} highlights the need to apply stroke-specific evaluation strategies. Therefore, for each of the segmented signal windows that contain a complete stroke motion sequence, our analysis starts with a stroke classification algorithm. We consider that a complete stroke motion sequence can be decomposed into distinct sequential phases\cite{anand2017wearable}, where the motion difference between phases is reflected in arm rotation, moving speed, and acceleration. Therefore, to reduce computation complexity and increase the feature reliability, we focus on IMU sensor data for the classification task. To enable robust stroke classification across users and reduce hand-crafted rule bias, we adopt a data-driven approach.

\subsubsection{Data Pre-processing and Feature Extraction}

For the 6-axis IMU signal from the windows, we first removed 20 frames (200ms in sample rate of 100Hz) from the head and the tail, respectively, to reduce boundary noise. The signals from each \revise{axis} were filtered using a second-order low-pass filter with a 20Hz cutoff to preserve human arm motion dynamics. 

In our motion classification tasks, designing an optimal feature extraction strategy is crucial to capture representative global motion pattern features\cite{gong2017pyro}. To this end, we extracted both time-domain and frequency-domain features from each axis of the signals for classification. We compute basic statistical descriptors and temporal features (e.g., sum, mean, variance, Standard deviation, skewness, kurtosis, percentiles, peak count, etc) on the raw signal and their $1^{st}$ derivative. We also extract features from the frequency domain by applying Fast Fourier Transform (FFT) to calculate spectral energy and statistical descriptors. We also applied Welch’s method for estimating the power spectral density, and computed its max and mean as a feature. As a result, we produced a $X \in \mathbb{R}^{6 \times 23}$ feature per stroke for the classification.

\subsubsection{Evaluation}

\begin{figure}[]
    \centering
    \includegraphics[width=0.65\linewidth]{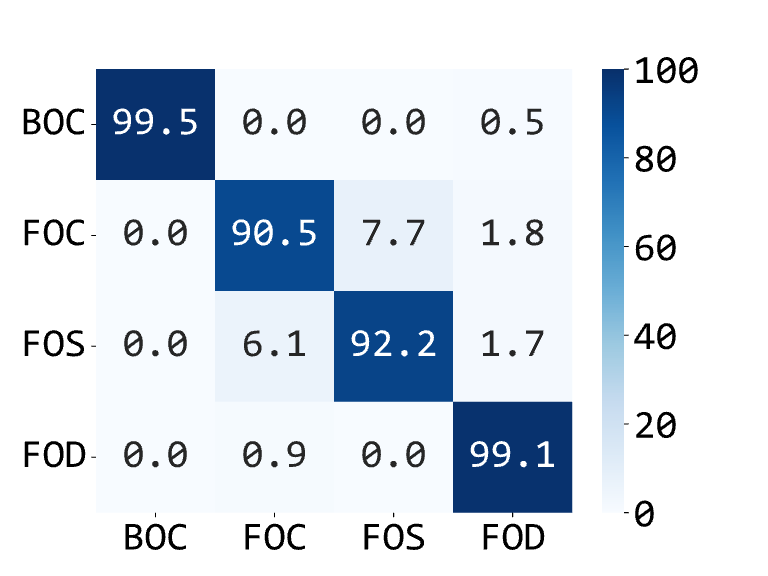}
    \caption{The confusion matrix of the SVM results. BOC, FOC, FOS, and FOD denote \textit{backhand overhead clear}, \textit{forehand overhead clear}, \textit{forehand overhead smash}, and \textit{forehand overhead drop}, respectively.}
    \label{fig:svm_cm} 
\end{figure}

To identify the optimal model for our task, we trained and evaluated multiple machine learning models on our dataset, including Support Vector Machine (SVM), Linear Regression (LinearReg), Random Forest (RF), and Multilayer Perceptron (MLP), under different training strategies. We first evaluate on a general model performance using a 5-fold cross-validation \revise{strategy}, by splitting \revise{the} data set into training and testing subsets in a 4:1 ratio, regardless of the user factor. We iteratively train and test with different partitions, and average the results over rounds. The results show that SVM achieves a \revise{highest} classification accuracy of 95.05\%. To eliminate the user-specific bias and test the model in a general real-world usage scenario, we adopt a leave-3-user-out training strategy, where we randomly pick the data from 9 users for training and test on the remaining 3 users. The results show that SVM also \revise{outperforms} the other \revise{classifiers}, achieving a recognition accuracy of 91.43\%. Figure. \ref{fig:svm_cm} shows the confusion matrix of SVM's performance. We noticed that \textit{forehand overhead clear} are prone to \revise{being misclassified} as \textit{forehand overhead smash} as they show highly similar patterns, especially in the first two phases (Figure. \ref{fig:coordinate}, \textcircled{2} and \textcircled{4}). As a result, we use this SVM model in the final implementation. 
Table.\ref{TAB:classification} summarizes the performance of different models across different training strategies. 

\begin{table*}[]
\begin{tabular}{l||cccc|cccc}
\hline
 & \multicolumn{4}{c|}{General} & \multicolumn{4}{c}{User-Independent} \\ \cline{2-9} 
 & Accuracy & Percision & Recall & F1 score & Accuracy & Percision & Recall & F1 score \\ \hline
\textit{SVM} & 95.05\% & 95.39\% & 95.28\% & 95.23\% & 91.43\% & 92.03\% & 91.99\% & 92.00\% \\
\textit{Linear Regression} & 54.95\% & 47.84\% & 39.69\% & 41.88\% & 41.90\% & 40.53\% & 28.66\% & 31.74\% \\
\textit{Random Forest} & 79.72\% & 81.19\% & 80.47\% & 80.29\% & 66.19\% & 75.43\% & 67.62\% & 66.63\% \\
\textit{MLP} & 92.57\% & 93.03\% & 92.91\% & 92.88\% & 80.95\% & 82.47\% & 81.86\% & 82.00\% \\ \hline
\end{tabular}
\caption{Stroke classification performance across models and training strategies}
\label{TAB:classification}
\end{table*}

\subsection{Stroke Quality Rating}

The task of evaluating the stroke quality involves estimating a rating score \revise{for} each given stroke signal. Inspired by \textit{I1} from \textbf{DR1}, we implemented stroke-specific rating strategies by training stroke-specific models in a data-driven context. \textbf{F1} further highlighted that the muscle strength plays a critical role in executing a correct and powerful stroke. Those muscle strength patterns would reflect sequentially across phases of the entire motion sequence that can be captured by IMU measurements, such as rotation and acceleration. Therefore, the rating algorithm must effectively capture the temporal relationships between consecutive sensor data frames.

We initially experimented with two rating strategies. One is to treat each motion score as \revise{a} discrete label and apply a classification approach. However, our \revise{pilot} test on this did not show promising results. One possible reason is that different assessors may follow varying rating criteria. As discussed in Section \ref{SEC:stroke_quality_rating}, although the rating score \revise{has} good reliability across assessors in \revise{the} ``consistency'' definition, the reliability becomes lower when \revise{considering} a ``absolute agreement'' definition, indicating the bias of rating criteria across assessors. To eliminate this bias, we treated the motion score as \revise{a} continuous numerical value, normalized \revise{it} across assessors, and then applied a regression-based approach for the task.

\subsubsection{Data Resampling and Pre-processing}

We observed that IMU sensors are prone to \revise{producing} measurement saturation during high-speed movements and sudden impacts, which occur frequently in our badminton stroking case. This will also lead to uneven signal samplings during wireless data transition. Since the stroke rating task is sensitive to temporal relationships between consecutive sensor data frames, we adopted the resampling strategy suggested in prior work\cite{liu2024smartdampener}, by applying cubic spline interpolation to the sensor data. This increases the sample rate from 100Hz to 500Hz. Following the preprocessing procedure used in our classification task, we removed 100 frames (200ms at 500Hz) from both the beginning and the end of each sequence to reduce boundary noise
% , and applied a low-pass filter with a cut-off frequency of 20Hz to remove high-frequency noise. 
As a result, we obtained a data vector of $X \in \mathbb{R}^{6 \times 800}$ per stroke for input to the regression model.

\subsubsection{Models}

We design, train, and evaluate through several machine learning models on our dataset, including traditional machine learning models such as Support Vector Regression (SVR) and Linear Regression (LinearReg), as well as deep learning approaches such as CNN-LSTM, MS-TCN, and a Transformer-based regression model. Please see Appendix \ref{appendix:a} for more models details.

% \rerevise{The CNN-LSTM model combines a Convolutional Neural Network (CNN) to extract spatial features from the input signals with a Long Short-Term Memory (LSTM) network to capture temporal dependencies across time steps. We consider that a completed stroke motion sequence can be decomposed into five sequential phases \cite{anand2017wearable}. \revise{Consequently}, we equally divided each stroke into five temporal segments and fed them into five CNN models, respectively. This produced five feature vectors representing five motion phases. These vectors were then processed by LSTM blocks, followed by a fully connected layer to generate the regression output.

% The MS-TCN model is a widely used convolutional layer-based neural network for action segmentation\cite{farha2019ms}, which has also been adapted and proven effective for processing IMU sensor data\cite{park2024silent}. Since the original MS-TCN model is designed for sequential segmentation tasks, we modified the model architecture by adding a fully connected layer after the TCN model to obtain a regression output. 

% The transformer module\cite{vaswani2017attention} is designed with \revise{an} attention mechanism that effectively represents sequential data by considering both past and future frames, which \revise{is} theoretically suitable for our task. Our implementation is based on a variation designed for processing IMU data\cite{shavit2021boosting},  with a modified output layer using a fully-connected layer to a regression score.}

\subsubsection{Evaluation}

For evaluation, we adopt the same strategy as in the stroke classification task: (1) a general model trained with 5-fold cross-validation, and (2) a user-independent model trained with a leave-user-out approach. For both settings, we trained four separate models corresponding to the four stroke types and evaluated them individually. We measure model performance by calculating the score (from 1 to 5) mean absolute error (MAE). The complete results are reported in Table \ref{TAB:evaluation}. The results show that SVR outperforms the other approaches in both training schemes, with an average MAE across four strokes at 0.423 for \revise{the} general model and 0.438 for \revise{the} user-independent model. Therefore, we use SVR as our final choice.

% Similar to the evaluation process for our stroke classification task, we evaluate on a general model using 5-fold cross-validation and train a user-independent model through a leave-user-out approach. For each approach, we trained four separated models for the four strokes and evaluate them separately. We measure to model performance by calculating the score (from 1 to 5) mean absolute error (MAE). The full results are shown in Table. \ref{TAB:evaluation}. 

\begin{table*}[t]
\centering
\caption{MAE of stroke rating performance across models and training strategies. A lower MAE indicates better results. BOC, FOC, FOS, and FOD denote \textit{backhand overhead clear}, \textit{forehand overhead clear}, \textit{forehand overhead smash}, and \textit{forehand overhead drop}, respectively. }
\begin{tabular}{l||ccccc|ccccc}
\hline
\textbf{}          & \multicolumn{5}{c|}{General}                                      & \multicolumn{5}{c}{User-Independent}                                                                                             \\ \cline{2-11} 
\textit{}          & BOC & FOC   & FOS   & \multicolumn{1}{l|}{FOD}   & AVG   & \multicolumn{1}{l}{BOC} & \multicolumn{1}{l}{FOC} & \multicolumn{1}{l}{FOS} & \multicolumn{1}{l|}{FOD} & \multicolumn{1}{l}{AVG} \\ \hline
\textit{SVR}       & 0.420        & 0.404 & 0.413 & \multicolumn{1}{l|}{0.456} & 0.423 & 0.411                     & 0.426                     & 0.432                     & \multicolumn{1}{c|}{0.483} & 0.438                     \\
\textit{LinearReg} & 1.269        & 0.542 & 0.576 & \multicolumn{1}{l|}{0.605} & 0.748 & 0.757                     & 0.771                     & 0.553                     & \multicolumn{1}{c|}{0.703} & 0.696                     \\
\textit{CNN-LSTM}  & 0.574        & 0.534 & 0.651 & \multicolumn{1}{l|}{0.513} & 0.568 & 0.700                     & 0.727                     & 0.595                     & \multicolumn{1}{c|}{0.474} & 0.624                     \\
MS-TCN            & 0.773        & 0.574 & 0.661 & \multicolumn{1}{l|}{0.668} & 0.669 & 0.578                     & 0.484                     & 0.516                     & \multicolumn{1}{c|}{0.683} & 0.565                     \\
\textit{Transformer}        & 0.658        & 0.540 & 0.601 & \multicolumn{1}{l|}{0.560} & 0.590 & 0.802                     & 0.571                     & 0.669                     & \multicolumn{1}{c|}{0.763} & 0.701                     \\ \hline
\end{tabular}
% \caption{MAE of stroke rating performance across models and training strategies. A lower MAE indicates better results. BOC, FOC, FOS, and FOD denote \textit{backhand overhead clear}, \textit{forehand overhead clear}, \textit{forehand overhead smash}, and \textit{forehand overhead drop}, respectively. }
\label{TAB:evaluation}
\end{table*}

\subsection{Impact Location Estimation}

Estimating the shuttle impact location is crucial for enabling fine-grained stroke evaluation and providing stroke-specific feedback (\textbf{DR2}). To the best of our knowledge, there is no prior work dedicated to estimating the impact location for the shuttle, either using a racket-mounted sensor or wearable devices. While estimating the impact location indirectly through a smartwatch is challenging, \textit{I2} points out that humans can infer the stroke impact location based on arm vibration and impact sounds. This observation inspires us to explore the potential of sensing impact location indirectly through arm vibration and acoustic signal. Therefore, we examine two feature extraction strategies: IMU feature only (IMU), and combining IMU feature with acoustic feature (Acoustic+IMU). 

\subsubsection{Data Pre-processing and Feature Extraction}

Theoretically, different shuttle impact locations should result in amplitude and frequency differences on the IMU readings. \textbf{F5} highlights that a center impact typically produces strong and prolonged racket vibration with a soft sound, while a side impact results in weaker and stiffer vibration with a crisp sound. 
% Furthermore, side impact on the left or right side can induce phase differences along \hl{XXX} axis, as the grip position is centered. 
Considering that both amplitude and phase variations are more effectively represented in the frequency domain, we extract the feature by transforming the signal into its frequency domain. 

For the IMU signals, we followed a pre-processing pipeline similar to that we used in the stroke quality rating task. Specifically, we applied cubic spline interpolation to upsample the raw data from 100 Hz to 500 Hz in order to refine the temporal resolution. While the shuttle impact itself is instantaneous, the resulting vibrations can persist for a longer duration. Since we ensured that the impact event is always located at the center of each stroke sequence, we cropped a 200-ms window around the midpoint as our region of interest. To capture the temporal–frequency feature, we applied the Short-time Fourier transform (STFT) on each IMU axis, yielding a feature map of $X \in \mathbb{R}^{6 \times 9 \times 26}$ for each impact. For the corresponding audio signals within the signal segment of interest, we 
% first applied a low-pass filter with a cut-off frequency of \hl{XXX}Hz to suppress the wind motion noise. 
follow a common practice for acoustic feature extraction\cite{Chen2021} by applying STFT directly to the audio signal to produce a frequency feature map for each impact event.

\subsubsection{Models and Pipeline}

\begin{figure*}[htb]
    \centering
    \includegraphics[width=\linewidth]{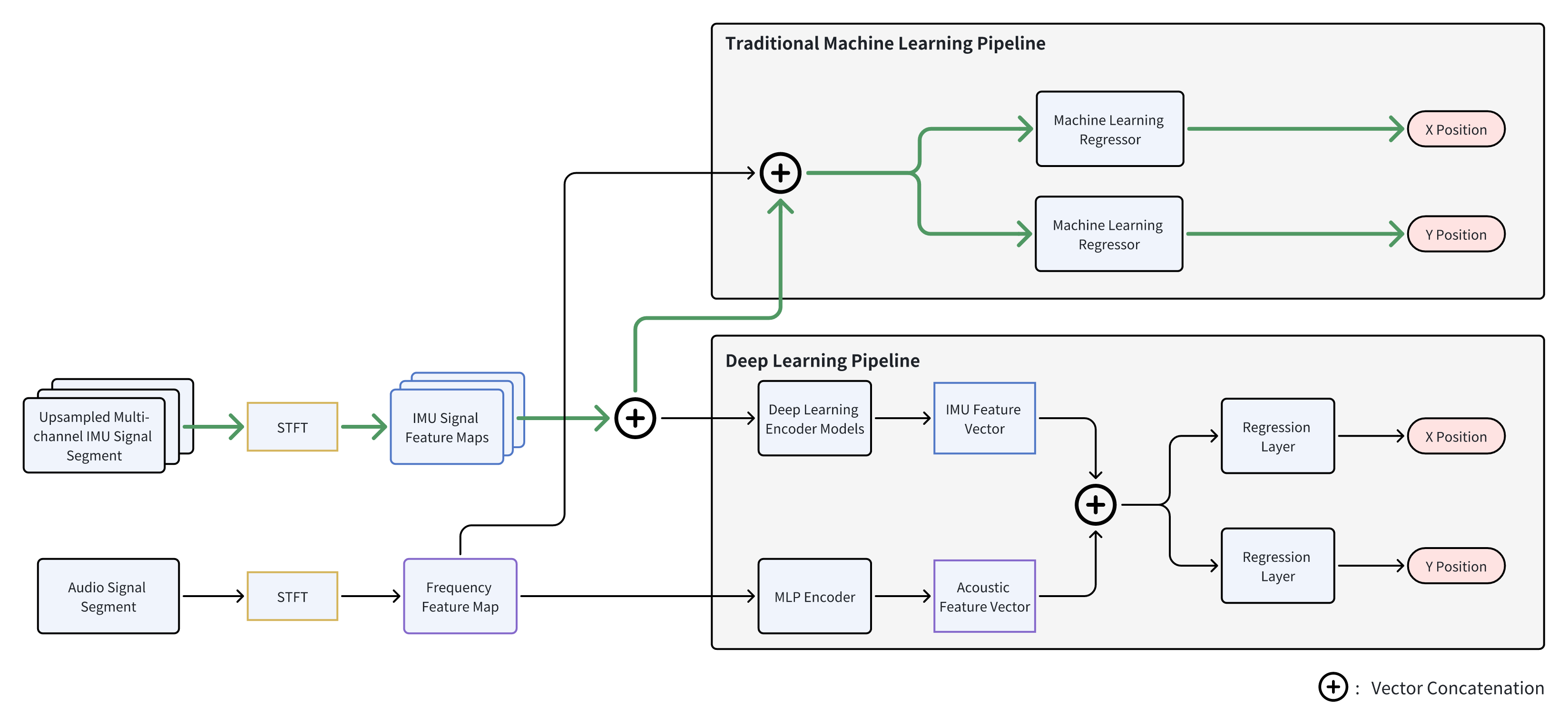}
    \caption{The multi-model signal processing pipeline for evaluating impact location estimation performance. We evaluate this pipeline by testing across multiple machine learning regressors and deep learning encoder models. The processing flow with a thick green line yields the base performance on our task, with an SVR model.}
    \label{fig:impact_location_pipeline} 
\end{figure*}

The impact location estimation task can also be formulated as a regression problem, where the impact position on the racket face is represented in a normalized Cartesian coordinate space. \revise{In particular, we set the throat of the racket as the origin. \revise{To ensure the range of both x and y axes is between 0 and 1, we normalize the x-axis to the interval [-0.5, 0.5], where -0.5 corresponds to the far left boundary on the racket face, and +0.5 corresponds to the far right boundary on it. The y-axis denotes the bottom-to-top dimension and normalizes to the interval of [0, 1], where 1 denotes the center of the upper edge (Figure. \ref{fig:impact_location_visualization}).}} We evaluated several regression models, including traditional machine learning models such as Support Vector Regression (SVR) and Linear Regression (LinearReg), and deep learning models such as CNN, MS-TCN, and a Transformer-based regressor. We initially experimented with two training strategies: (1) treating each location coordinate as a 2D vector and computing a joint regression loss, and (2) treating each axis independently by training two separate models. Our pilot test shows that using separate models yields more promising results, and thus, we adopted it in the subsequent evaluations.

Different from the stroke quality rating task, where we use a CNN-LSTM model to capture sequential phase dependencies, the impact location estimation relies only on short impact clips. Therefore, we employed a pure CNN model and an IMU data encoder without recurrent components. The main structure of the MS-TCN model and transformer model is identical to the previous experience, with a small change in the input layer to fit the STFT size. 

Figure. \ref{fig:impact_location_pipeline} illustrates the signal processing pipeline for this study. The IMU frequency feature map was concatenated before feeding it into the machine learning model or deep learning encoder models. The acoustic frequency feature map was concatenated with the IMU feature map directly for the machine learning model. For deep learning models, we introduce a simple MLP encoder to process the acoustic frequency feature map. And we adopt a late-fusion strategy to concatenate the acoustic feature vector with the IMU feature vector produced from the deep learning encoder models for the coordinates regression.

\subsubsection{Evaluation}

We train the models using 5-fold cross-validation and leave-user-out validation under two feature extraction strategies: IMU feature only (IMU), and combining IMU feature with acoustic feature (Acoustic+IMU), and evaluated across multiple machine learning models (Table. \ref{TAB:impact_location}).
 
\begin{figure*}[h]
    \centering
    \includegraphics[width=0.75\linewidth]{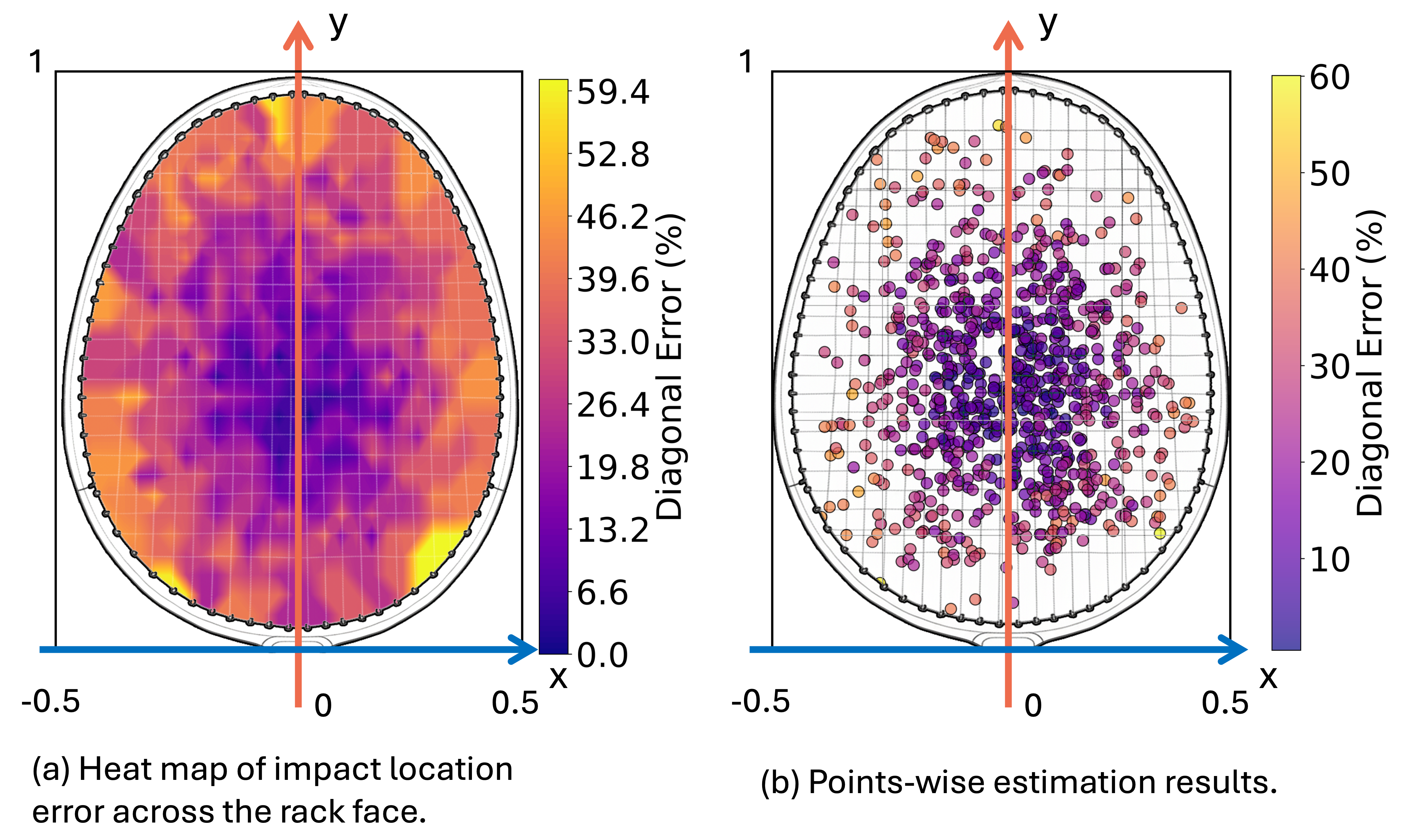}
    \caption{Visualization of the impact location estimation results.}
    \label{fig:impact_location_visualization} 
\end{figure*}

We measured the normalized MAE of the shuttle impact location along the X axis and the Y axis. Results show that SVR achieves the best overall performance with an average MAE of 0.132 in the general model, with both feature strategies yielding comparable results. Under the user-independent training scheme, which better reflects real-world performance, SVR with IMU features and CNN with IMU features both perform best, achieving an average MAE of 0.129. We observe a small advantage of the presence of acoustic features, but not significant, especially for superior models such as SVR and CNN. Calculating acoustic features also increases computational cost, resulting in a longer latency during real-world usage. Moreover, the SVR model generally inference faster than the CNN model. These findings and considerations motivate us to choose SVM with IMU feature in our final implementation. A heatmap of estimated impact locations across the racket face using the SVR model is shown in Figure. \ref{fig:impact_location_visualization}(a). We also visualize the individual prediction of each location point in Figure. \ref{fig:impact_location_visualization}(b). As shown in the figure, the closer to the edge of the racket face, the greater the error rate. One of the reasons is that most of the impact location in our dataset was located in the central area of the racket face (Figure. \ref{fig:impact_location_visualization}(b)), resulting in fewer training samples observed of edge cases. Moreover, impacts near the edge area typically generate a weaker vibration response, which makes them harder to distinguish from each other. Interestingly, we observed that the MAE along the Y axis is generally lower than the X axis across all models and feature settings. An explanation for this is that the gripping position is aligned with the racket's central axis, off-center impacts along the Y axis would induce larger rotational movement compared with the X axis, resulting in a higher estimation performance along the Y axis.

\begin{table*}[t]
\centering
\caption{The normalized MAE of the impact location estimate performance on X and Y axis across models, feature selection strategies, and training strategies. A lower MAE indicates better results.}
\begin{tabular}{l||l|ccc|ccc}
\hline
\multirow{2}{*}{} & \multicolumn{1}{c|}{\multirow{2}{*}{Feature}} & \multicolumn{3}{c|}{General} & \multicolumn{3}{c}{User-Independent} \\ \cline{3-8} 
 & \multicolumn{1}{c|}{} & X & \multicolumn{1}{c|}{Y} & AVG & X & \multicolumn{1}{c|}{Y} & AVG \\ \hline
\multirow{2}{*}{\textit{SVR}} & IMU & 0.136 & \multicolumn{1}{c|}{0.129} & \textbf{0.132} & 0.133 & \multicolumn{1}{c|}{0.124} &  \textbf{0.129}\\
 & Acoustic+IMU & \multicolumn{1}{l}{0.135} & \multicolumn{1}{l|}{0.130} & \multicolumn{1}{l|}{\textbf{0.132} } & \multicolumn{1}{l}{0.135} & \multicolumn{1}{l|}{0.133} & \multicolumn{1}{l}{\textbf{0.134}} \\ \hline
\multirow{2}{*}{\textit{LinearReg}} & IMU & 0.346 & \multicolumn{1}{c|}{0.320} & 0.333 & 0.336 & \multicolumn{1}{c|}{0.263} & 0.300 \\
 & Acoustic+IMU & \multicolumn{1}{l}{0.168} & \multicolumn{1}{l|}{0.159} & \multicolumn{1}{l|}{0.164} & \multicolumn{1}{l}{0.162} & \multicolumn{1}{l|}{0.173} & \multicolumn{1}{l}{0.168} \\ \hline
\multirow{2}{*}{\textit{CNN}} & IMU & 0.136 & \multicolumn{1}{c|}{0.134} & 0.135 & 0.134 & \multicolumn{1}{c|}{0.124} & \textbf{0.129} \\
 & Acoustic+IMU & \multicolumn{1}{l}{0.138} & \multicolumn{1}{l|}{0.143} & \multicolumn{1}{l|}{0.140} & \multicolumn{1}{l}{0.136} & \multicolumn{1}{l|}{0.138} & \multicolumn{1}{l}{0.137} \\ \hline
\multirow{2}{*}{\textit{MS-TCN}} & IMU & 0.135 & \multicolumn{1}{c|}{0.137} & 0.136 & 0.134 & \multicolumn{1}{c|}{0.126} & 0.130 \\
 & Acoustic+IMU & \multicolumn{1}{l}{0.135} & \multicolumn{1}{l|}{0.140} & \multicolumn{1}{l|}{0.137} & \multicolumn{1}{l}{0.134} & \multicolumn{1}{l|}{0.137} & \multicolumn{1}{l}{0.135} \\ \hline
\multirow{2}{*}{\textit{Transformer}} & IMU & 0.136 & \multicolumn{1}{c|}{0.139} & 0.137 & 0.144 & \multicolumn{1}{c|}{0.126} & 0.135 \\
 & Acoustic+IMU & \multicolumn{1}{l}{0.135} & \multicolumn{1}{l|}{0.139} & \multicolumn{1}{l|}{0.137} & \multicolumn{1}{l}{0.142} & \multicolumn{1}{l|}{0.149} & \multicolumn{1}{l}{0.146} \\ \hline
\end{tabular}
% \caption{The normalized MAE of the impact location estimate performance on X and Y axis across models, feature selection strategies, and training strategies. A lower MAE indicates better results.}
\label{TAB:impact_location}
\end{table*}

\subsection{Implementation}

The final implementation of \sysName~ consists of a smartwatch front-end application, a Python-based back-end server, and a Progressive Web App (PWA) based on the React framework to visualize the results. The smartwatch was developed using the WearOS framework, based on V Mollyn et al.'s implementation\cite{mollyn2022samosa}, and we deployed it on a Samsung Galaxy Watch 6 FE smartwatch. The back-end server was developed based on the Scikit-Learn library, and we deployed it on a Windows Laptop PC with an 11th Gen Intel(R) Core(TM) i5-11300H CPU and 16 GB of RAM. 

We implemented a rule-based algorithm to generate the improvement advice based on the statistical analysis results, incorporating the insights from our post-session interview (Figure. \ref{fig:optimal_impact_location}) (\textbf{DR3}). 

\revise{
\subsection{System Workflow and User Interaction}

We demonstrate the system workflow of BadminSense through a running example. Alan is a badminton enthusiast who wants to track and monitor his on-court badminton performance. Before starting, he connects the smartwatch to Wi-Fi to access the Internet. He launches the BadminSense App on the smartwatch and waits for the system to establish a server connection. Once connected, he starts the tracking by tapping the start button on the smartwatch and begins his badminton session. After finishing the session, he taps the stop button on the smartwatch to trigger the data upload and analysis process. Approximately one minute later, the smartwatch sends a notification indicating that the analysis is completed and displays a QR code to show the detailed results. Alan scans the QR Code with his smartphone to access the mobile interface, and he can view the session statistic from three main pages (Figure. \ref{fig:teaser}). 
% to view the details. The mobile interface consists of three main pages (Fig. \ref{fig:teaser}) for this session: 
% \begin{enumerate}
% [label={(\arabic*)}, leftmargin=*]
% \item Session Details: This page shows the overall statistics of this session, including total stroke count, average stroking performance, and breakdown performance for each type of stroke. 
% \item Session Timeline: This page displays snapshots of all strokes from this session in chronological order on a scrolling page. 
% \item Stroke Details: This page presents fine-grained information for each individual stroke, including stroke type, time stamp, stroke rating, visualization of the impact location, and improvement advice text. 
% \end{enumerate}
Besides, Alan can browse all past session information for long-term performance tracking. The mobile interface also provides an overall statistic page that reports the cumulative stroke count and average performance scores across all past sessions. It also calculates the proportion and average rating for each stroke type to provide him with an overview of his strengths and weaknesses.
}

\section{User Evaluation}

We conducted a usability study of \sysName~ to evaluate its performance in a real-world badminton practice scenario, where we aim to understand user satisfaction when practicing with the system. Given the fact that there are no devices or systems currently providing similar features to \sysName, it is hard to conduct a pair-wise comparison with a baseline system in terms of the system performance and the system usability. Therefore, our study is focusing on users' subjective feedback of the system from three aspects: 1) the reliability of the system performance, including the precision of stroke rating, impact location, and the improvement feedback, 2) the usability of the system, and 3) their engagement and reward during the experience with our system.

\subsection{Participants}

We recruited 12 participants (10 male, 2 female) from the institutional badminton club for the study. Our recruitment follows a criterion of requiring participants to have a minimum badminton experience of 2 years to ensure they have sufficient expertise to provide valuable feedback on the system performance. This also guarantees that participants were capable of self-assessing their own performance and provided a reliable evaluation of \sysName's~ performance in comparison with their self-assessment. \revise{The study was reviewed and approved by the institutional Human Research Ethics Committee. All participants were provided informed consent prior to participation.}

\subsection{Procedure}

Prior to the study, participants completed a demographic questionnaire. After a brief introduction of the study purpose, each participant was asked to wear a Samsung Galaxy Watch on their dominant wrist, with the option to choose between a 42mm or 48mm model to fit their wrist. Participants were then asked to play a single tournament game (21 points in total) using a racket strung at 24 lbs tension. To simulate a real-world usage scenario, we did not restrict the participants to perform a specific stroke during the tournament. Instead, we asked them to play the game as they normally would. After the game, we presented their performance results through the \sysName~ interface and explained how to interpret the results, where we particularly focused on stroke rating, impact location, and feedback advice. Finally, participants completed a post-study questionnaire to assess their perceptions of the system. \revise{The study lasted around 15 minutes, and each participant received a 3 USD coupon for compensation.}

\subsection{Questionnaire Design}

We carefully designed a post-study questionnaire with 5-point Likert scale items that cover the following three factors:

\begin{enumerate}[label={}, leftmargin=*]

\item \textbf{Performance.} This section evaluated the reliability of system performance. These items are designed to align the fine-grained statistical stroke evaluation outputs (i.e, stroke classification, stroke rating, impact location estimation, and feedback advice) with participants’ expectations and their perceived usefulness for skill improvement (Q1–Q8).

\item \textbf{System Usability.} This section evaluated the overall usability of \sysName. These items were carefully adapted and refined from the standard System Usability Scale (SUS) to better fit the context of \sysName~(Q9–Q14).

\item \textbf{User Engagement.} This section measured participants’ engagement throughout the experience. In particular, we were interested in whether the presence of \sysName~ disrupted the natural gaming flow. We also pay particular attention to users' perceptions of the system's value. These items were carefully adapted from the standard User Engagement Scale (UES-SF) - Short Form\cite{OBRIEN201828} to better fit the context of \sysName~(Q15–Q22).

\end{enumerate}

\subsection{Results}

% \begin{figure*}[htb]
%     \centering
%     \includegraphics[width=\linewidth]{figs/SUS.png}
%     \caption{The detailed questionnaire results distribution from the user evaluation. The digit on the bar indicates the number of ratings. }
%     \label{fig:sus} 
% \end{figure*}

Participants received an average stroke rating of 2.9. The highest score was 3.8, achieved by P6, an experienced player with approximately 5 years of \revise{playing} experiences \revise{and 2 years of coaching experiences}. His teammates, P7 and P8, gained the second and third-highest average scores at 3.6 and 3.5, respectively. The lowest score was 2.4, recorded by P2, who had around two years of experience. However, since we did not explicitly require participants to perform only our four focused strokes, the rating results contain bias as the motion sequence should involve the other types of strokes, which \sysName~ is not yet designed to handle at this stage. 

For the questionnaire results, we reverse the scores of Q10, Q15, and Q18-Q20 before calculating the statistics of each factor. The results indicate that participants consistently felt satisfied with and benefited from \sysName's~ feedback $(M = 4.14, SD = 0.17)$. They also agree that the system has good usability $(M = 4.27, SD = 0.21)$ and reported feeling engaged while using \sysName~$(M = 4.24, SD = 0.30)$. These findings suggest that \sysName~ was able to provide useful and motivating performance feedback without disturbing users' natural badminton gaming flow. Please see Appendix \ref{appendix:b} for more details.

P6 mentioned that he had prior experience using a racket-handle-mounted sensor for badminton, and he commented that \sysName~ is \textit{flexible and convenient} and could provide \textit{professional metric} compared to other solutions. P1 (21 years old, 5 years of experience) was impressed by the stroke quality rating function, and commented that \sysName~ has a \textit{user-friendly interface} with \textit{high portability}. P3 (19 years old, 2 years of experience) felt \sysName~ is \textit{useful for novice user}, but he needs more time with it to use to build trust in the results. P11 (32 years old, 3 years of experience) particularly appreciates the impact of location detection results, and comments that \textit{``I cannot wait to use it for practicing my stroke''}.

% 教练
% 见过放在拍尾的传感器，他认为这个比较便捷，不需要额外携带设备
% 同学 
% 惊讶，可以检测并分辨动作，虽然是新手，但仍然具有一定识别精度。
% 对新手友好，系统操作简单，只需要一个手表，便捷。
% 用户界面简洁明了，可以快速找到需要的功能。
% 如果足够准确，认为对新手训练是有帮助的，但还需要一段时间的体验才能确信。
% 路人
% 练习时，可能用来调整击球点

\revise{

\section{Discussion}

\subsection{Effect of Stroke Side on System Performance}

Overall, our system performs consistently across varying stroke types with different stroke sides (i.e., forehand stroke and backhand stroke) for both stroke-quality rating and impact-location estimation. However, we observed slight performance drops for backhand shots (BOC) compared to the other forehand shots. For stroke quality rating, BOC shows a higher mean error relative to FOC, FOS, and FOD, especially on the user-independent model (Table. \ref{TAB:evaluation}). A similar pattern appears in impact location estimation, where BOC yields an MAE of 0.146, compared with an average MAE of 0.125 across all forehand strokes. We attribute this outcome to two factors. First, as shown in Figure. \ref{fig:coordinate}(b), backhand strokes(BOC) generally produce less significant inertial measurements compared to those of a forehand stroke, which makes them more difficult for the model to extract meaningful features. Second, our dataset size of backhand strokes is smaller than that of forehand strokes. While this imbalance reflects the stroke type frequency in real-world badminton play \cite{phomsoupha2015science}, it would also limit the machine learning models' generalizability across all types of strokes.

\subsection{Limited Contribution of Acoustic Features on Impact Location Estimation}

Inspired by \textbf{F5} that indicates badminton players often assess their stroke quality by the impact sound, we initially incorporate the acoustic signal as a feature for the impact location estimation. However, our results suggest that incorporating acoustic features offers limited benefit for impact location estimation and may even degrade model performance in some cases (Table. \ref{TAB:impact_location}). Our explanation for this outcome is that both the IMU signals and the acoustic signals capture vibration-related features. However, airborne sound propagation in a noisy environment is likely to introduce signal attenuation, making the acoustic signal less reliable than an IMU that captures the same information through bone propagation. As a result, the use of acoustic signals not only fails to provide more meaningful features but also introduces additional noise, yielding noticeable performance decline. While \textit{LinearReg} shows improvement when acoustic signals are included, it is the only linear model that cannot effectively leverage the richer features from the IMU. The presence of the acoustic signal provides additional amplitude information, which increases its performance, but it still lags behind other models that rely solely on IMU features.

\subsection{How Users Interpret and Benefit From \sysName's Feedback}

To investigate how users interpret the results from \sysName~ and how these results benefit their badminton training process, we conducted interviews with participants from the user evaluation and summarized several key observations. Stroke type classification results helped participants identify their weak stroke types (\textbf{F2}). Participants consistently mention that, combined with the post-stroke performance quality scores, they become more aware of their strengths and weaknesses in different stroke types, and could adjust their practice focus accordingly. It also served as positive reinforcement when high scores were achieved. The impact location results provide intuitive and quantified information about the shuttle's impact area, enabling more precise adjustments than relying solely on subjective vibration or sound cues (\textbf{F4}, \textbf{F5}). Participants also appreciated the improvement advice. They noted that beginners often lack guidance from professional coaches during their daily training, and can only improve their skills through intuition or by comparing their moves with other players. The timely improvement advice acts as an in-situ virtual coach to facilitate beginners' early development (\textbf{F1}). Lastly, participants mentioned that session-level recordings are useful for comparing and visualizing cross-match performance (\textbf{F3}). Based on this, they can adjust their strategies for future matches or training sessions to enhance their performance. Overall, participants at different skill levels agree that \sysName's performance metrics, such as stroke type, stroke quality rating, impact location, and improvement advice, are helpful for improving their skill.

\subsection{Supporting Deliberate Practice with \sysName}

Participants also envision using \sysName~ to support structured deliberate practice. One participant specifically inquired how the system could provide immediate stroke-by-stroke feedback to support more structured stroke training, much like receiving in-suit coaching. In the current implementation, participants can create a one-stroke session by manually starting and stopping the tracking session before and after each stroke to obtain performance analysis. While feasible, participants comment that this process is imaginable tedious and disruptive to their natural practice pace. Their suggestions highlight a potential and valuable direction for improvement: developing a "Deliberate Practice Mode" that automatically detects a stroke and outputs instant feedback without manual intervention. This feature is technically feasible with our current pipeline and represents one of the promising directions for our future improvement. 

}

\section{Limitation and Future Works}

In this paper, although we showed the potential of \sysName~ for providing fine-grained badminton performance analysis and demonstrated its feasibility in real-world usage scenarios, there are still some limitations and potential for future improvements.

\subsection{Supporting Broader Stroke Types}

As a proof-of-concept, \sysName~ currently focuses on a subset of common strokes in badminton. Although this selection covers the most common actions in amateur play, supporting a broader badminton stroke is necessary to enable more general and professional usage cases. One simple and straightforward way to achieve this is to expand the dataset by collecting data samples of other types of strokes and retraining the system. However, we also observed that some of the badminton strokes share similar motion across multiple phases. For example, the motion sequence first two phases of \textit{forehand overhead clear} and \textit{forehand overhead smash} shows highly similar patterns (Figure. \ref{fig:coordinate}, \textcircled{2} and \textcircled{4}). This highlights the potential to adapt few-shot or transfer learning approaches\cite{xu2022enabling} from existing datasets, thereby reducing the burden of large-scale data collection.

\subsection{Effects on Racket String Tension}

As a proof-of-concept, \sysName~ currently is trained and developed based on a dataset collected using a badminton racket with a string tension of 24lbs, as we consider that this is the most common and adaptive setting that balances shuttle controllability and hitting power. However, players with different skill levels often prefer different string tension settings: looser strings provide more power but with lower controllability, whereas tighter strings enhance control but reduce power. Such tension variation may affect sensor reading, particularly for impact location estimation, where the vibration and acoustic characteristics generated from the string area may vary. The future work should focus on examining the effect of string tension on our method, or seeking calibration mechanisms to adapt the algorithm to individual racket attributes.

\subsection{Handedness Issue}

Currently, the development and evaluation of \sysName~ only involve right-handed players. While the string tension variation primarily influences the impact location estimation, the handedness potentially affects the stroke classification and quality rating results, as the movement patterns of left-handed players differ from those of right-handed players. As future work, it is crucial to explore mechanisms to support left-handed user usage. As suggested in previous research\cite{liu2024smartdampener}, one potential approach is to train the algorithm with an augmented dataset where the sensor data axis is swapped and mirrored to simulate data from left-hand movements.

\subsection{Speed Estimation}

Another limitation of \sysName~ is that we did not explicitly estimate the shuttlecock speed. \textbf{F4} revealed that participants agree that the speed and direction of the shuttle are closely related to stroke motion quality and impact location, which motivated us to focus on analyzing player behaviors rather than the shuttle dynamics. However, it could be argued that estimating shuttle speed may serve as a more direct indicator of stroke quality, but speed estimation remains a challenging problem. Prior work in tennis\cite{liu2024smartdampener} has attempted to use physical models for speed estimation, but they found that this is less reliable compared to data-driven methods. Alternatively, another research focusing on baseball\cite{lopez2019site} employed a data-driven approach by collecting the ground truth of baseball speed with radar speed guns. However, badminton presents unique challenges: the shuttlecock is both lightweight and hollow, which may be hard to capture with a speed gun, and its rapid flying speed makes reliable data collection far more difficult. As future work, it is worth investigating more effective and efficient methods to estimate the shuttle speed, potentially by combining physical modeling with data-driven approaches to balance between feasibility and accuracy.

\subsection{Toward Context-Aware Improvement Advice}

Currently, the generation of improvement advice feedback is rule-based, referring to our preliminary interview with experienced players. While this approach is effective for providing basic feedback as a proof-of-concept, it has limited flexibility. One limitation is that the current advice did not consider historical performance. As future work, we plan to explore more comprehensive and context-aware feedback mechanisms based on the fine-grained statistics produced by our method. One feasible and promising approach is incorporating large language models (LLMs) to generate context-aware coaching advice, considering the player's individual historical performance trends and training goals. 

\revise{
\subsection{System Reliability and Extendability}

Our current usability evaluation focuses on short-term performance by collecting user-perceived reliability. While participants reported strong alignment between \sysName's output and their expectations, such assessments were relatively subjective. However, it is unfeasible to conduct real-world quantity evaluation since this would require expert stroke-by-stroke quality rating and impact location annotation, which would disrupt the natural badminton activity flow. One promising way for evaluating \sysName's quantitative performance is examining the users' skill improvement through a longitudinal study, which will be our future work. Furthermore, the sensing pipeline of \sysName~ only relies on acoustic and IMU data. As MEMS microphones and IMU can be embedded in diverse wrist-worn wearable devices such as smart bracelets and smart wristbands, we believe that our dataset and sensing approach can be easily adapted to other form factors. As future work, we will explore and evaluate extending \sysName~ to other wrist-worn devices. 
}

\section{Conclusion}

In this paper, we presented \sysName, a badminton stroke evaluation system that leverages off-the-shelf smartwatch sensing to provide fine-grained feedback on players’ stroke quality rating and impact location. The design of \sysName~ was informed by two formative interviews with experienced players, which helped us derive design requirements and implementation insights. We then conduct a data collection user study to collect data with fine-grained labels, including stroke type, expert-assessed stroke rating, and shuttle impact location. We trained and evaluated \sysName~ through a series of experiments to evaluate its offline performance. Finally, we conducted a usability study that demonstrated \sysName's feasibility in real-world usage and highlighted its potential to provide unobtrusive and meaningful support for daily badminton training. We further envision the future extensions of \sysName, such as supporting more stroke types, adapting to diverse racket attributes, estimating shuttle speed, and generating context-aware coaching feedback. \sysName~ highlights the broader potential of wrist-worn devices to enable fine-grained sensing and analysis of racket sport, which could be further extended to other sports involving swinging motions, such as golf, tennis, and table tennis.

\begin{acks}

This research was partially supported by the National
Natural Science Foundation of China, Young Scientists Fund (Project No. 62402301), Natural Science Foundation of Guangdong Province, General Research Fund (Project No. 2025A1515010236), the STU Scientific Research Initiation Grant (SRIG, Project No. NTF23024), and Scientific Development Funds from Shenzhen University (Project No. 000001032518). 

\end{acks}

\bibliographystyle{ACM-Reference-Format}
\bibliography{sample-base}

@String{Computing = "Computing" }

@String{Computer = "{IEEE} Computer" }

@String{Springer = "Springer-Verlag" }

@article{reid2007skill,
  title={Skill acquisition in tennis: Research and current practice},
  author={Reid, Machar and Crespo, Miguel and Lay, Brendan and Berry, Jason},
  journal={Journal of science and medicine in sport},
  volume={10},
  number={1},
  pages={1--10},
  year={2007},
  publisher={Elsevier},
  doi={10.1016/j.jsams.2006.05.011}
}

@inproceedings{elvitigala2024grand,
  title={Grand Challenges in SportsHCI},
  author={Elvitigala, Don Samitha and Karahano{\u{g}}lu, Arma{\u{g}}an and Matviienko, Andrii and Turmo Vidal, Laia and Postma, Dees and Jones, Michael D and Montoya, Maria F and Harrison, Daniel and Elb{\ae}k, Lars and Daiber, Florian and others},
  booktitle={Proceedings of the 2024 CHI Conference on Human Factors in Computing Systems},
  pages={1--20},
  year={2024},
  doi={10.1145/3613904.3642050}
}

@article{phomsoupha2015science,
  title={The science of badminton: game characteristics, anthropometry, physiology, visual fitness and biomechanics},
  author={Phomsoupha, Michael and Laffaye, Guillaume},
  journal={Sports medicine},
  volume={45},
  number={4},
  pages={473--495},
  year={2015},
  publisher={Springer},
  doi={https://doi.org/10.1007/s40279-014-0287-2}
}

@article{secckin2023review,
  title={Review on Wearable Technology in sports: Concepts, Challenges and opportunities},
  author={Se{\c{c}}kin, Ahmet {\c{C}}a{\u{g}}da{\c{s}} and Ate{\c{s}}, Bahar and Se{\c{c}}kin, Mine},
  journal={Applied sciences},
  volume={13},
  number={18},
  pages={10399},
  year={2023},
  publisher={MDPI},
  doi={https://doi.org/10.3390/app131810399}
}

@article{aroganam2019review,
  title={Review on wearable technology sensors used in consumer sport applications},
  author={Aroganam, Gobinath and Manivannan, Nadarajah and Harrison, David},
  journal={Sensors},
  volume={19},
  number={9},
  pages={1983},
  year={2019},
  publisher={MDPI},
  doi={https://doi.org/10.3390/s19091983}
}

@inproceedings{turmo2021design,
  title={The design space of wearables for sports and fitness practices},
  author={Turmo Vidal, Laia and Zhu, Hui and Waern, Annika and M{\'a}rquez Segura, Elena},
  booktitle={Proceedings of the 2021 CHI Conference on Human Factors in Computing Systems},
  pages={1--14},
  year={2021},
  doi={https://doi.org/10.1145/3411764.3445700}
}

@article{liu2024smartdampener,
  title={SmartDampener: An Open Source Platform for Sport Analytics in Tennis},
  author={Liu, Runze and Lu, Taiting and Yuan, Shengming and Zhou, Hao and Gowda, Mahanth},
  journal={Proceedings of the ACM on Interactive, Mobile, Wearable and Ubiquitous Technologies},
  volume={8},
  number={3},
  pages={1--30},
  year={2024},
  publisher={ACM New York, NY, USA},
  doi={https://doi.org/10.1145/3678507}
}

@online{sony,
	title = {Sony Smart Tennis Sensor},
	url = {http://tennis-technology.com/sony-smart-tennis-sensor/},
	titleaddon = {Tennis-Technology},
	urldate = {2025-07-08},
	langid = {american},
    year = {2015}
}

@online{qlipp_tennis_qlipp_2017,
	title = {Qlipp Tennis},
	url = {https://www.eedesignit.com/the-tennis-sensor-thats-making-a-racket/},
	titleaddon = {{eeDesignIt}.com},
	author = {Qlipp Tennis},
	urldate = {2025-07-08},
	date = {2017-04-17},
	langid = {american},
}

@online{babolat_play,
	title = {Babolat Play Review},
	url = {https://tennis-technology.com/babolat-play-review/},
	titleaddon = {Tennis-Technology},
	urldate = {2025-07-18},
	langid = {american},
    year = {2015}
}

@online{zepp,
	title = {Zepp Tennis Sensor Review},
	url = {https://tennis-technology.com/zepp-tennis-sensor-review/},
	titleaddon = {Tennis-Technology},
	urldate = {2025-07-18},
	langid = {american},
    year = {2015}
}

@online{swingvision,
	title = {{SwingVision}: {AI} Scoring, Stats \& Line Calling},
	url = {https://swing.vision/},
	shorttitle = {{SwingVision}},
	urldate = {2025-07-07},
	langid = {english},
    year = {2025}
}

@online{playsight,
	title = {playsight},
	url = {https://playsight.com},
	urldate = {2025-07-08},
	date = {2020-08-05},
	langid = {american},
    year = {2020}
}

@online{hawk-eye,
	title = {Hawk-Eye {\textbar} A global leader in the live sports arena},
	url = {https://www.hawkeyeinnovations.com/www.hawkeyeinnovations.com/},
	urldate = {2025-07-08},
	date = {2025-06-04},
	langid = {english},
}

@inproceedings{park2024silent,
  title={Silent Impact: Tracking Tennis Shots from the Passive Arm},
  author={Park, Junyong and Yang, Saelyne and Jo, Sungho},
  booktitle={Proceedings of the 37th Annual ACM Symposium on User Interface Software and Technology},
  pages={1--15},
  year={2024},
  doi={https://doi.org/10.1145/3654777.3676403}
}

@inproceedings{anand2017wearable,
  title={Wearable motion sensor based analysis of swing sports},
  author={Anand, Akash and Sharma, Manish and Srivastava, Rupika and Kaligounder, Lakshmi and Prakash, Divya},
  booktitle={2017 16th IEEE international conference on machine learning and applications (ICMLA)},
  pages={261--267},
  year={2017},
  organization={IEEE},
  doi={10.1109/ICMLA.2017.0-149}
}

@online{vivo,
	title = {vivo {WATCH} {GT} {\textbar} vivo Hong Kong},
	url = {https://www.vivo.com/hk/en/products/watch-gt},
	urldate = {2025-07-18},
    year={2025},
}

@online{oppo,
	title = {{OPPO} Watch X {\textbar} {OPPO} United Kingdom},
	url = {https://www.oppo.com/uk/accessories/watch-x/},
	titleaddon = {{OPPO}},
	urldate = {2025-07-18},
	langid = {british},
    year={2025},
}

@inproceedings{coorevits2016rise,
  title={The rise and fall of wearable fitness trackers},
  author={Coorevits, Lynn and Coenen, Tanguy},
  booktitle={Academy of Management},
  year={2016},
  doi={https://doi.org/10.5465/AMBPP.2016.17305abstract}
}

@online{balance,
	title = {Balance 2},
	url = {https://us.amazfit.com/products/balance-2},
	titleaddon = {Amazfit},
	urldate = {2025-07-18},
	langid = {english},
    year={2025},
}

@inproceedings{ibh2024stroke,
  title={A stroke of genius: Predicting the next move in badminton},
  author={Ibh, Magnus and Gra{\ss}hof, Stella and Hansen, Dan Witzner},
  booktitle={Proceedings of the IEEE/CVF Conference on Computer Vision and Pattern Recognition},
  pages={3376--3385},
  year={2024},
  doi={https://doi.org/10.1109/CVPRW63382.2024.00342}
}

@inproceedings{anik2016activity,
  title={Activity recognition of a badminton game through accelerometer and gyroscope},
  author={Anik, Md Ariful Islam and Hassan, Mehedi and Mahmud, Hasan and Hasan, Md Kamrul},
  booktitle={2016 19th International Conference on Computer and Information Technology (ICCIT)},
  pages={213--217},
  year={2016},
  organization={IEEE},
  doi={https://doi.org/10.1109/ICCITECHN.2016.7860197}
}

@inproceedings{weeratunga2017application,
  title={Application of computer vision and vector space model for tactical movement classification in badminton},
  author={Weeratunga, Kokum and Dharmaratne, Anuja and Boon How, Khoo},
  booktitle={Proceedings of the IEEE conference on computer vision and pattern recognition workshops},
  pages={76--82},
  year={2017},
  doi={https://doi.org/10.1109/CVPRW.2017.22}
}

@inproceedings{weeratunga2014application,
  title={Application of computer vision to automate notation for tactical analysis of badminton},
  author={Weeratunga, Kokum and How, Khoo Boon and Dharmaratne, Anuja and Messom, Chris},
  booktitle={2014 13th International Conference on Control Automation Robotics \& Vision (ICARCV)},
  pages={340--345},
  year={2014},
  organization={IEEE},
  doi={https://doi.org/10.1109/ICARCV.2014.7064329}
}

@article{steels2020badminton,
  title={Badminton activity recognition using accelerometer data},
  author={Steels, Tim and Van Herbruggen, Ben and Fontaine, Jaron and De Pessemier, Toon and Plets, David and De Poorter, Eli},
  journal={Sensors},
  volume={20},
  number={17},
  pages={4685},
  year={2020},
  publisher={MDPI},
  doi={https://doi.org/10.3390/s20174685}
}

@inproceedings{peralta2022badminton,
  title={Badminton stroke classification based on accelerometer data: from individual to generalized models},
  author={Peralta, Daniel and Van Herbruggen, Ben and Fontaine, Jaron and Debyser, Wout and Wieme, Jorg and De Poorter, Eli},
  booktitle={2022 IEEE International Conference on Big Data (Big Data)},
  pages={5542--5548},
  year={2022},
  organization={IEEE},
  doi={https://doi.org/10.1109/BigData55660.2022.10020984}
}

@article{wang2016badminton,
  title={Badminton stroke recognition based on body sensor networks},
  author={Wang, Zhelong and Guo, Ming and Zhao, Cong},
  journal={IEEE Transactions on Human-Machine Systems},
  volume={46},
  number={5},
  pages={769--775},
  year={2016},
  publisher={IEEE},
  doi={https://doi.org/10.1109/THMS.2016.2571265}
}

@inproceedings{chu2017badminton,
  title={Badminton video analysis based on spatiotemporal and stroke features},
  author={Chu, Wei-Ta and Situmeang, Samuel},
  booktitle={Proceedings of the 2017 ACM on international conference on multimedia retrieval},
  pages={448--451},
  year={2017},
  doi={https://doi.org/10.1145/3078971.3079032}
}

@article{wang2023benchmarking,
  title={Benchmarking stroke forecasting with stroke-level badminton dataset},
  author={Wang, Wei-Yao and Du, Wei-Wei and Peng, Wen-Chih and Ik, Tsi-Ui},
  journal={arXiv preprint arXiv:2306.15664},
  year={2023},
  doi={https://doi.org/10.48550/arXiv.2306.15664}
}

@article{chang2025bst,
  title={BST: Badminton Stroke-type Transformer for Skeleton-based Action Recognition in Racket Sports},
  author={Chang, Jing-Yuan},
  journal={arXiv preprint arXiv:2502.21085},
  year={2025},
  doi={https://doi.org/10.48550/arXiv.2502.21085}
}

@article{towler2023effects,
  title={Effects of racket moment of inertia on racket head speed, impact location and shuttlecock speed during the badminton smash},
  author={Towler, Harley and Mitchell, SR and King, MA},
  journal={Scientific Reports},
  volume={13},
  number={1},
  pages={14060},
  year={2023},
  publisher={Nature Publishing Group UK London},
  doi={https://doi.org/10.1038/s41598-023-37108-x}
}

@article{van2024strategy,
  title={Strategy analysis of badminton players using deep learning from IMU and UWB wearables},
  author={Van Herbruggen, Ben and Fontaine, Jaron and Simoen, Jonas and De Mey, Lennert and Peralta, Daniel and Shahid, Adnan and De Poorter, Eli},
  journal={Internet of Things},
  volume={27},
  pages={101260},
  year={2024},
  publisher={Elsevier},
  doi={https://doi.org/10.1016/j.iot.2024.101260}
}

@article{zhu2025analysis,
  title={The analysis of motion recognition model for badminton player movements using machine learning},
  author={Zhu, Xuanmin and Liu, Lizhi and Huang, Jingshuo and Chen, Genyan and Ling, Xi and Chen, Yanshuo},
  journal={Scientific Reports},
  volume={15},
  number={1},
  pages={1--17},
  year={2025},
  publisher={Nature Publishing Group},
  doi={https://doi.org/10.1038/s41598-025-02771-9}
}

@article{lin2023effect,
  title={The effect of wearable technology on badminton learning performance: a multiple feedback WISER model in physical education},
  author={Lin, Kuo-Chin and Hung, Hui-Chun and Chen, Nian-Shing},
  journal={Smart Learning Environments},
  volume={10},
  number={1},
  pages={28},
  year={2023},
  publisher={Springer},
  doi={https://doi.org/10.1186/s40561-023-00247-9}
}

@inproceedings{rahmad2020vision,
  title={Vision based automated badminton action recognition using the new local convolutional neural network extractor},
  author={Rahmad, Nur Azmina and As’ ari, Muhammad Amir and Ibrahim, Mohamad Fauzi and Sufri, Nur Anis Jasmin and Rangasamy, Keerthana},
  booktitle={Enhancing Health and Sports Performance by Design: Proceedings of the 2019 Movement, Health \& Exercise (MoHE) and International Sports Science Conference (ISSC)},
  pages={290--298},
  year={2020},
  organization={Springer},
  doi={https://doi.org/10.1007/978-981-15-3270-2_30}
}

@inproceedings{wang2019automatic,
  title={Automatic badminton action recognition using cnn with adaptive feature extraction on sensor data},
  author={Wang, Ya and Fang, Weichuang and Ma, Jinwen and Li, Xiangchen and Zhong, Albert},
  booktitle={International Conference on Intelligent Computing},
  pages={131--143},
  year={2019},
  organization={Springer},
  doi={https://doi.org/10.1007/978-3-030-26763-6_13}
}

@inproceedings{blank2017ball,
  title={Ball speed and spin estimation in table tennis using a racket-mounted inertial sensor},
  author={Blank, Peter and Groh, Benjamin H and Eskofier, Bjoern M},
  booktitle={Proceedings of the 2017 ACM International Symposium on Wearable Computers},
  pages={2--9},
  year={2017},
  doi={https://doi.org/10.1145/3123021.3123040}
}

@inproceedings{blank2015sensor,
  title={Sensor-based stroke detection and stroke type classification in table tennis},
  author={Blank, Peter and Ho{\ss}bach, Julian and Schuldhaus, Dominik and Eskofier, Bjoern M},
  booktitle={Proceedings of the 2015 ACM International Symposium on Wearable Computers},
  pages={93--100},
  year={2015},
  doi={https://doi.org/10.1145/2802083.2802087}
}

@inproceedings{diogo2025smashing,
  title={Smashing Insights: Prototyping a Video-Based System For Racket Sports},
  author={Diogo, Jo{\~a}o and Rodrigues, Rui and Martins, Tom{\'a}s and Correia, Nuno},
  booktitle={Proceedings of the Extended Abstracts of the CHI Conference on Human Factors in Computing Systems},
  pages={1--10},
  year={2025},
  doi={https://doi.org/10.1145/3706599.3719865}
}

@inproceedings{zhao2019tenniseye,
  title={TennisEye: tennis ball speed estimation using a racket-mounted motion sensor},
  author={Zhao, Hongyang and Wang, Shuangquan and Zhou, Gang and Jung, Woosub},
  booktitle={Proceedings of the 18th international Conference on information Processing in Sensor Networks},
  pages={241--252},
  year={2019},
  doi={https://doi.org/10.1145/3302506.3310404}
}

@inproceedings{yang2017tennismaster,
  title={TennisMaster: An IMU-based online serve performance evaluation system},
  author={Yang, Disheng and Tang, Jian and Huang, Yang and Xu, Chao and Li, Jinyang and Hu, Liang and Shen, Guobin and Liang, Chieh-Jan Mike and Liu, Hengchang},
  booktitle={Proceedings of the 8th augmented human international conference},
  pages={1--8},
  year={2017},
  doi={https://doi.org/10.1145/3041164.3041186}
}

@inproceedings{gourgari2013thetis,
  title={Thetis: Three dimensional tennis shots a human action dataset},
  author={Gourgari, Sofia and Goudelis, Georgios and Karpouzis, Konstantinos and Kollias, Stefanos},
  booktitle={Proceedings of the IEEE conference on computer vision and pattern recognition workshops},
  pages={676--681},
  year={2013},
  doi={https://doi.org/10.1109/CVPRW.2013.102}
}

@inproceedings{lopez2019site,
  title={On-site personal sport skill improvement support using only a smartwatch},
  author={Lopez, Guillaume and Abe, Shohei and Hashimoto, Kengo and Yokokubo, Anna},
  booktitle={2019 IEEE International Conference on Pervasive Computing and Communications Workshops (PerCom Workshops)},
  pages={158--164},
  year={2019},
  organization={IEEE},
  doi={10.1109/PERCOMW.2019.8730681}
}

@article{ganser2021classification,
  title={Classification of tennis shots with a neural network approach},
  author={Ganser, Andreas and Hollaus, Bernhard and Stabinger, Sebastian},
  journal={Sensors},
  volume={21},
  number={17},
  pages={5703},
  year={2021},
  publisher={MDPI},
  doi={https://doi.org/10.3390/s21175703}
}

@article{jia2021swingnet,
  title={SwingNet: Ubiquitous fine-grained swing tracking framework via stochastic neural architecture search and adversarial learning},
  author={Jia, Hong and Hu, Jiawei and Hu, Wen},
  journal={Proceedings of the ACM on Interactive, Mobile, Wearable and Ubiquitous Technologies},
  volume={5},
  number={3},
  pages={1--21},
  year={2021},
  publisher={ACM New York, NY, USA},
  doi={https://doi.org/10.1145/3478082}
}

@article{sharma2018wearable,
  title={Wearable audio and IMU based shot detection in racquet sports},
  author={Sharma, Manish and Anand, Akash and Srivastava, Rupika and Kaligounder, Lakshmi},
  journal={arXiv preprint arXiv:1805.05456},
  year={2018},
  doi={https://doi.org/10.48550/arXiv.1805.05456}
}

@article{skublewska-paszkowska_temporal_2023,
	title = {Temporal Pattern Attention for Multivariate Time Series of Tennis Strokes Classification},
	volume = {23},
	rights = {https://creativecommons.org/licenses/by/4.0/},
	issn = {1424-8220},
	url = {https://www.mdpi.com/1424-8220/23/5/2422},
	doi = {10.3390/s23052422},
	pages = {2422},
	number = {5},
	journaltitle = {Sensors},
	author = {Skublewska-Paszkowska, Maria and Powroznik, Pawel},
	urldate = {2025-07-25},
	date = {2023-02-22},
	langid = {english},
	note = {Publisher: {MDPI} {AG}}
}

@inproceedings{ma2024avattar,
  title={avaTTAR: Table Tennis Stroke Training with Embodied and Detached Visualization in Augmented Reality},
  author={Ma, Dizhi and Hu, Xiyun and Shi, Jingyu and Patel, Mayank and Jain, Rahul and Liu, Ziyi and Zhu, Zhengzhe and Ramani, Karthik},
  booktitle={Proceedings of the 37th Annual ACM Symposium on User Interface Software and Technology},
  pages={1--16},
  year={2024},
  doi={https://doi.org/10.1145/3654777.3676400}
}

@inproceedings{weng2025bridging,
  title={Bridging Coaching Knowledge and AI Feedback to Enhance Motor Learning in Basketball Shooting Mechanics Through a Knowledge-Based SOP Framework},
  author={Weng, Jian-Jia and Ku, Calvin and Wang, Jo Chien and Cheng, Chih-Jen and Lin, Tica and Su, Yu-An and Tsai, Tsung-Hsun and Lin, You-Yi and Ku, Lun-Wei and Chu, Hung-Kuo and others},
  booktitle={Proceedings of the 2025 CHI Conference on Human Factors in Computing Systems},
  pages={1--20},
  year={2025},
  doi={https://doi.org/10.1145/3706598.3713324}
}

@inproceedings{zhu2023iball,
  title={iball: Augmenting basketball videos with gaze-moderated embedded visualizations},
  author={Zhu-Tian, Chen and Yang, Qisen and Shan, Jiarui and Lin, Tica and Beyer, Johanna and Xia, Haijun and Pfister, Hanspeter},
  booktitle={Proceedings of the 2023 CHI Conference on Human Factors in Computing Systems},
  pages={1--18},
  year={2023},
  doi={https://doi.org/10.1145/3544548.3581266}
}

@article{liu2022posecoach,
  title={Posecoach: A customizable analysis and visualization system for video-based running coaching},
  author={Liu, Jingyuan and Saquib, Nazmus and Zhutian, Chen and Kazi, Rubaiat Habib and Wei, Li-Yi and Fu, Hongbo and Tai, Chiew-Lan},
  journal={IEEE Transactions on Visualization and Computer Graphics},
  volume={30},
  number={7},
  pages={3180--3195},
  year={2022},
  publisher={IEEE},
  doi={10.1109/TVCG.2022.3230855}
}

@article{lin2022quest,
  title={The quest for omnioculars: Embedded visualization for augmenting basketball game viewing experiences},
  author={Lin, Tica and Zhu-Tian, Chen and Yang, Yalong and Chiappalupi, Daniele and Beyer, Johanna and Pfister, Hanspeter},
  journal={IEEE transactions on visualization and computer graphics},
  volume={29},
  number={1},
  pages={962--972},
  year={2022},
  publisher={IEEE},
  doi={10.1109/TVCG.2022.3209353}
}

@inproceedings{cheng2024viscourt,
  title={VisCourt: In-Situ Guidance for Interactive Tactic Training in Mixed Reality},
  author={Cheng, Liqi and Jia, Hanze and Yu, Lingyun and Wu, Yihong and Ye, Shuainan and Deng, Dazhen and Zhang, Hui and Xie, Xiao and Wu, Yingcai},
  booktitle={Proceedings of the 37th Annual ACM Symposium on User Interface Software and Technology},
  pages={1--14},
  year={2024},
  doi={https://doi.org/10.1145/3654777.3676466}
}

@inproceedings{wu2023ar,
  title={Ar-enhanced workouts: Exploring visual cues for at-home workout videos in ar environment},
  author={Wu, Yihong and Yu, Lingyun and Xu, Jie and Deng, Dazhen and Wang, Jiachen and Xie, Xiao and Zhang, Hui and Wu, Yingcai},
  booktitle={Proceedings of the 36th annual ACM symposium on user interface software and technology},
  pages={1--15},
  year={2023},
  doi={https://doi.org/10.1145/3586183.3606796}
}

@article{ericsson1993role,
  title={The role of deliberate practice in the acquisition of expert performance.},
  author={Ericsson, K Anders and Krampe, Ralf T and Tesch-R{\"o}mer, Clemens},
  journal={Psychological review},
  volume={100},
  number={3},
  pages={363},
  year={1993},
  publisher={American Psychological Association},
  doi={https://doi.org/10.1037//0033-295X.100.3.363}
}

@article{zhang2012effects,
  title={Effects of play practice on teaching table tennis skills},
  author={Zhang, Peng and Ward, Phillip and Li, Weidong and Sutherland, Sue and Goodway, Jackie},
  journal={Journal of Teaching in Physical Education},
  volume={31},
  number={1},
  pages={71--85},
  year={2012},
  publisher={Human Kinetics, Inc.},
  doi={https://doi.org/10.1123/jtpe.31.1.71}
}

@article{arbabi2016effect,
  title={Effect of performance feedback with three different video modeling methods on acquisition and retention of badminton long service},
  author={Arbabi, Ahmad and Sarabandi, Maliheh},
  journal={Sport Science},
  volume={9},
  pages={41--45},
  year={2016},
  doi={}
}

@article{krizkova2021sport,
  title={Sport performance analysis with a focus on racket sports: A review},
  author={Krizkova, Sarka and Tomaskova, Hana and Tirkolaee, Erfan Babaee},
  journal={Applied Sciences},
  volume={11},
  number={19},
  pages={9212},
  year={2021},
  publisher={MDPI},
  doi={https://doi.org/10.3390/app11199212}
}

@article{laffaye2015changes,
  title={Changes in the game characteristics of a badminton match: a longitudinal study through the olympic game finals analysis in men’s singles},
  author={Laffaye, Guillaume and Phomsoupha, Michael and Dor, Fr{\'e}d{\'e}ric},
  journal={Journal of sports science \& medicine},
  volume={14},
  number={3},
  pages={584},
  year={2015},
  doi={}
}

@inproceedings{huang2002kinematic,
  title={Kinematic analysis of three different badminton backhand overhead strokes},
  author={Huang, Kuei-Shu and Huang, Chenfu and Chung, Shaw Shiun and Tsai, Chien-Lu},
  booktitle={ISBS-Conference Proceedings Archive},
  year={2002},
  doi={https://doi.org/10.13140/2.1.4284.2881}
}

@book{grice_badminton_2008,
	title = {Badminton: Steps to Success},
	isbn = {978-0-7360-7229-8},
	url = {https://books.google.com.hk/books?id=Es5JngEACAAJ},
	series = {Sports instruction series},
	publisher = {Human Kinetics},
	author = {Grice, T.},
	date = {2008},
	lccn = {2007032476},
}

@article{mcerlain2020effect,
  title={Effect of racket-shuttlecock impact location on shot outcome for badminton smashes by elite players},
  author={McErlain-Naylor, Stuart A and Towler, Harley and Afzal, Idrees A and Felton, Paul J and Hiley, Michael J and King, Mark A},
  journal={Journal of sports sciences},
  volume={38},
  number={21},
  pages={2471--2478},
  year={2020},
  publisher={Taylor \& Francis},
  doi={https://doi.org/10.1080/02640414.2020.1792132}
}

@article{koo2016guideline,
  title={A guideline of selecting and reporting intraclass correlation coefficients for reliability research},
  author={Koo, Terry K and Li, Mae Y},
  journal={Journal of chiropractic medicine},
  volume={15},
  number={2},
  pages={155--163},
  year={2016},
  publisher={Elsevier},
  doi={https://doi.org/10.1016/j.jcm.2016.02.012}
}

@inproceedings{gong2017pyro,
  title={Pyro: Thumb-tip gesture recognition using pyroelectric infrared sensing},
  author={Gong, Jun and Zhang, Yang and Zhou, Xia and Yang, Xing-Dong},
  booktitle={Proceedings of the 30th Annual ACM Symposium on User Interface Software and Technology},
  pages={553--563},
  year={2017},
  doi={https://doi.org/10.1145/3126594.3126615}
}

@inproceedings{farha2019ms,
  title={Ms-tcn: Multi-stage temporal convolutional network for action segmentation},
  author={Farha, Yazan Abu and Gall, Jurgen},
  booktitle={Proceedings of the IEEE/CVF conference on computer vision and pattern recognition},
  pages={3575--3584},
  year={2019},
  doi={https://doi.org/10.1109/CVPR.2019.00369}
}

@article{vaswani2017attention,
  title={Attention is all you need},
  author={Vaswani, Ashish and Shazeer, Noam and Parmar, Niki and Uszkoreit, Jakob and Jones, Llion and Gomez, Aidan N and Kaiser, {\L}ukasz and Polosukhin, Illia},
  journal={Advances in neural information processing systems},
  volume={30},
  year={2017},
  doi={https://doi.org/10.48550/arXiv.1706.03762}
}

@article{shavit2021boosting,
  title={Boosting inertial-based human activity recognition with transformers},
  author={Shavit, Yoli and Klein, Itzik},
  journal={IEEE Access},
  volume={9},
  pages={53540--53547},
  year={2021},
  publisher={IEEE},
  doi={10.1109/ACCESS.2021.3070646}
}

@article{Chen2021,
    title = {{GestOnHMD : Enabling Gesture-based Interaction on Low-cost VR Head-Mounted Display}},
    year = {2021},
    journal = {IEEE Transactions on Visualization and Computer Graphics},
    author = {Chen, Taizhou and Xu, Lantian and Xu, Xianshan and Kening, Zhu},
    number = {5},
    pages = {2597--2607},
    volume = {27},
    doi = {10.1109/TVCG.2021.3067689}
}

@inproceedings{xu2022enabling,
  title={Enabling hand gesture customization on wrist-worn devices},
  author={Xu, Xuhai and Gong, Jun and Brum, Carolina and Liang, Lilian and Suh, Bongsoo and Gupta, Shivam Kumar and Agarwal, Yash and Lindsey, Laurence and Kang, Runchang and Shahsavari, Behrooz and others},
  booktitle={Proceedings of the 2022 CHI Conference on Human Factors in Computing Systems},
  pages={1--19},
  year={2022},
  doi={https://doi.org/10.1145/3491102.3501904}
}

@article{OBRIEN201828,
title = {A practical approach to measuring user engagement with the refined user engagement scale (UES) and new UES short form},
journal = {International Journal of Human-Computer Studies},
volume = {112},
pages = {28-39},
year = {2018},
issn = {1071-5819},
doi = {https://doi.org/10.1016/j.ijhcs.2018.01.004},
url = {https://www.sciencedirect.com/science/article/pii/S1071581918300041},
author = {Heather L. O’Brien and Paul Cairns and Mark Hall}
}

@article{mollyn2022samosa,
  title={SAMoSA: Sensing activities with motion and subsampled audio},
  author={Mollyn, Vimal and Ahuja, Karan and Verma, Dhruv and Harrison, Chris and Goel, Mayank},
  journal={Proceedings of the ACM on Interactive, Mobile, Wearable and Ubiquitous Technologies},
  volume={6},
  number={3},
  pages={1--19},
  year={2022},
  publisher={ACM New York, NY, USA},
  doi={https://doi.org/10.1145/3550284}
}

@article{lin2023vird,
  title={VIRD: immersive match video analysis for high-performance badminton coaching},
  author={Lin, Tica and Aouididi, Alexandre and Zhu-Tian, Chen and Beyer, Johanna and Pfister, Hanspeter and Wang, Jui-Hsien},
  journal={IEEE transactions on visualization and computer graphics},
  volume={30},
  number={1},
  pages={458--468},
  year={2023},
  publisher={IEEE},
  doi={10.1109/TVCG.2023.3327161}
}

@article{chu2021tivee,
  title={TIVEE: Visual exploration and explanation of badminton tactics in immersive visualizations},
  author={Chu, Xiangtong and Xie, Xiao and Ye, Shuainan and Lu, Haolin and Xiao, Hongguang and Yuan, Zeqing and Zhu-Tian, Chen and Zhang, Hui and Wu, Yingcai},
  journal={IEEE Transactions on Visualization and Computer Graphics},
  volume={28},
  number={1},
  pages={118--128},
  year={2021},
  publisher={IEEE},
  doi={10.1109/TVCG.2021.3114861}
}

@article{ye2020shuttlespace,
  title={Shuttlespace: Exploring and analyzing movement trajectory in immersive visualization},
  author={Ye, Shuainan and Zhu-Tian, Chen and Chu, Xiangtong and Wang, Yifan and Fu, Siwei and Shen, Lejun and Zhou, Kun and Wu, Yingcai},
  journal={IEEE transactions on visualization and computer graphics},
  volume={27},
  number={2},
  pages={860--869},
  year={2020},
  publisher={IEEE},
  doi={10.1109/TVCG.2020.3030392}
}

\appendix

\begin{figure*}[h]
    \centering
    \includegraphics[width=0.95\linewidth]{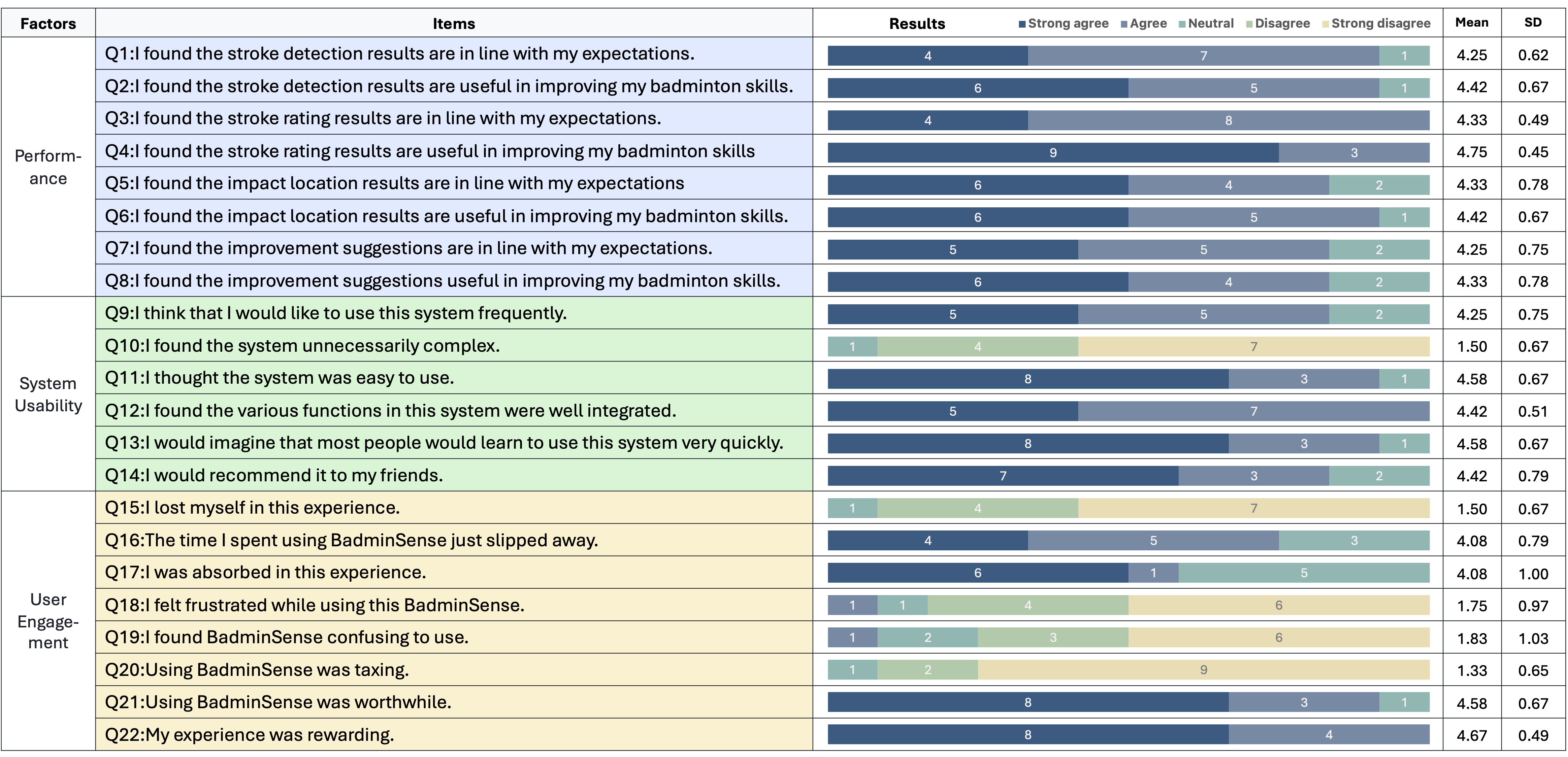}
    \caption{The detailed questionnaire results distribution from the user evaluation. The digit on the bar indicates the number of ratings. }
    \label{fig:sus} 
\end{figure*}

\section{Model Architecture for Stroke Quality Rating}
\label{appendix:a}

The CNN-LSTM model combines a Convolutional Neural Network (CNN) to extract spatial features from the input signals with a Long Short-Term Memory (LSTM) network to capture temporal dependencies across time steps. We consider that a completed stroke motion sequence can be decomposed into five sequential phases \cite{anand2017wearable}. \revise{Consequently}, we equally divided each stroke into five temporal segments and fed them into five CNN models, respectively. This produced five feature vectors representing five motion phases. These vectors were then processed by LSTM blocks, followed by a fully connected layer to generate the regression output.

The MS-TCN model is a widely used convolutional layer-based neural network for action segmentation\cite{farha2019ms}, which has also been adapted and proven effective for processing IMU sensor data\cite{park2024silent}. Since the original MS-TCN model is designed for sequential segmentation tasks, we modified the model architecture by adding a fully connected layer after the TCN model to obtain a regression output.

The transformer module\cite{vaswani2017attention} is designed with \revise{an} attention mechanism that effectively represents sequential data by considering both past and future frames, which \revise{is} theoretically suitable for our task. Our implementation is based on a variation designed for processing IMU data\cite{shavit2021boosting},  with a modified output layer using a fully-connected layer to a regression score.

\section{Questionnaire Items and Results}
\label{appendix:b}

Figure \ref{fig:sus} shows the detailed questionnaire items and results of our user evaluation. Each question can be rated from 0 to 5, of which 0 is strong disagree and 5 is strong agree. The digit on the bar indicates the number of ratings.

% \appendix









\end{document}